\documentclass[preprint,12pt]{elsarticle}
\usepackage[top=1in, bottom=1in, left=0.95in, right=0.95in]{geometry}

\usepackage{graphicx}
\usepackage{amssymb}
\usepackage{amsthm}
\usepackage{amsmath}

\usepackage{lineno}
\usepackage{subfigure}

\usepackage[utf8]{inputenc}
\usepackage{tabu}
\usepackage{xcolor}

\makeatletter
\def\ps@pprintTitle{%
   \let\@oddhead\@empty
   \let\@evenhead\@empty
   \let\@oddfoot\@empty
   \let\@evenfoot\@oddfoot
}
\makeatother

\begin{document}

\begin{frontmatter}

\title{Machine learning modeling of high entropy alloy: the role of short-range order \footnote{\footnotesize{This manuscript has been co-authored by UT-Battelle, LLC, under contract DE-AC05-00OR22725 with the US Department of Energy (DOE). The US government retains and the publisher, by accepting the article for publication, acknowledges that the US government retains a nonexclusive, paid-up, irrevocable, worldwide license to publish or reproduce the published form of this manuscript, or allow others to do so, for US government purposes. DOE will provide public access to these results of federally sponsored research in accordance with the DOE Public Access Plan (http://energy.gov/downloads/doe-public-access-plan).}}}

\author{Xianglin Liu \corref{cor1}$^\dagger$}
\ead{xianglinliu01@gmail.com}
\address{Materials Science and Technology Division, Oak Ridge National Laboratory}

\author{Jiaxin Zhang \corref{cor2} $^\dagger$ \footnote{\footnotesize{ $^\dagger$ These two authors contributed equally to this work}}}
\ead{jiaxin.zhanguq@gmail.com}
\address{Center for Computational Sciences, Oak Ridge National Laboratory}

\author{Markus Eisenbach}
\address{Center for Computational Sciences, Oak Ridge National Laboratory}

\author{Yang Wang}
\address{Pittsburgh Supercomputing Center, Carnegie Mellon University}

\begin{abstract}
The development of machine learning sheds new light on the traditionally complicated problem of thermodynamics in multicomponent alloys. Successful application of such a method, however, strongly depends on the quality of the data and model. Here we propose a scheme to improve the representativeness of the data by utilizing the short-range order (SRO) parameters to survey the configuration space. Using the improved data, a pair interaction model is trained for the NbMoTaW high entropy alloy using linear regression. Benefiting from the physics incorporated into the model, the learned effective Hamiltonian demonstrates excellent predictability over the whole configuration space. By including pair interactions within the 6th nearest-neighbor shell, this model achieves an $R^2$ testing score of 0.997 and root mean square error of 0.43 meV. We further perform a detailed analysis on the effects of training data, testing data, and model parameters. The results reveal the vital importance of representative data and physical model. On the other hand, we also examined the performance neural networks, which is found to demonstrate a strong tendency to overfit the data.

\end{abstract}

\begin{keyword}
first-principle calculation \sep machine learning \sep high entropy alloys \sep short-range order
\end{keyword}

\end{frontmatter}


\section{Introduction}

The density functional theory (DFT) provides a powerful method to calculate the properties of materials from first principles. While essentially a ground state theory, the DFT method can be applied to study materials at finite temperature when combined with Monte Carlo simulation. While the synergy of DFT and Monte Carlo is simple in principle, their application is severely hindered by the daunting computational cost. For example, using a supercell of 250-atom \cite{PhysRevB.93.024203}, the ``brute-force" calculation of the order-disorder transition in CuZn alloy, which is one of the earliest alloys used by humans, already takes $10^{8}$ CPU hours. One less ambitious, but more practical approach, is to establish an effective Hamiltonian from the DFT data, and feed this easy-to-calculate fitted model into MC simulations. One good example of this strategy is the cluster expansion method \cite{PhysRev.81.988, SANCHEZ1984334}, in which the energy is expressed in terms of the cluster functions and the corresponding effective cluster interactions (ECIs) \cite{RubanECIReview}. To determine the ECIs, the Connolly and Williams approach (structure inversion method) is typically applied, where the energies of $N_b$ different structures are calculated with the DFT method. While the cluster expansion provides a systematic way to construct the effective Hamiltonian, its application to multiple component systems is limited due to the rapid increase of the number of ECIs with respect to chemical components \cite{PhysRevLett.116.105501, seko2009cluster}. This is particularly true for the high entropy alloys (HEAs) \cite{ADEM:ADEM200300567, CANTOR2004213}, which are a class of metallic materials that contain more than 4 different principal elements, and demonstrate some exceptional mechanical properties \cite{NatureComNiCoCr, Gludovatz1153, SENKOV2011698, MIRACLE2017448}.

The progress of computing power and algorithms brings new opportunities to the modeling of thermodynamics in multicomponent systems \cite{Eisenbach_2019}. On one hand, with enhanced computational capability, a much larger set of DFT data can be calculated to determine the effective Hamiltonian. On the other hand, novel optimization techniques such as genetic algorithms\cite{PhysRevB.72.165113, doi:10.1021/ja9105623}, Bayesian approaches \cite{PhysRevB.80.024103, PhysRevB.81.012104}, and machine learning  \cite{ML_Cluster, Korman_npj} are developed for the modeling of complex systems. The machine learning method, in particular, has attracted significant attention due to its huge success on many fields \cite{NatureDL}, and has already been widely applied to the modeling of various physical quantities, such as atomic forces \cite{Chmielae1603015, PhysRevLett.114.096405}, interatomic potentials \cite{PhysRevB.95.094203, PhysRevX.8.041048, 2015arXiv151209110G}, and formation energies \cite{ML_Cluster, Ye2018}. Compared to other fields, application of the machine learning to the thermodynamics of HEAs faces additional challenges originated from the huge configuration space. As a result, for a robust machine learning model, the following requirements need to be satisfied:
\begin{itemize}
\item
The data should be representative. This demands spreading a sufficiently large amount of data over the configuration space to take into account all important features.
\item
The model must capture the underlying physics. This is the key for the model to maintain excellent reliability and predictability so that the Monte Carlo simulation can safely visit regions of the configuration space that are not well sampled by the training data. 
\item 
The testing data should be truly independent of the training data to avoid a spuriously high prediction accuracy due to the correlation between training and testing data. 
\end{itemize}

\begin{figure}[h!]
\centering
   \includegraphics[width=0.9\textwidth]{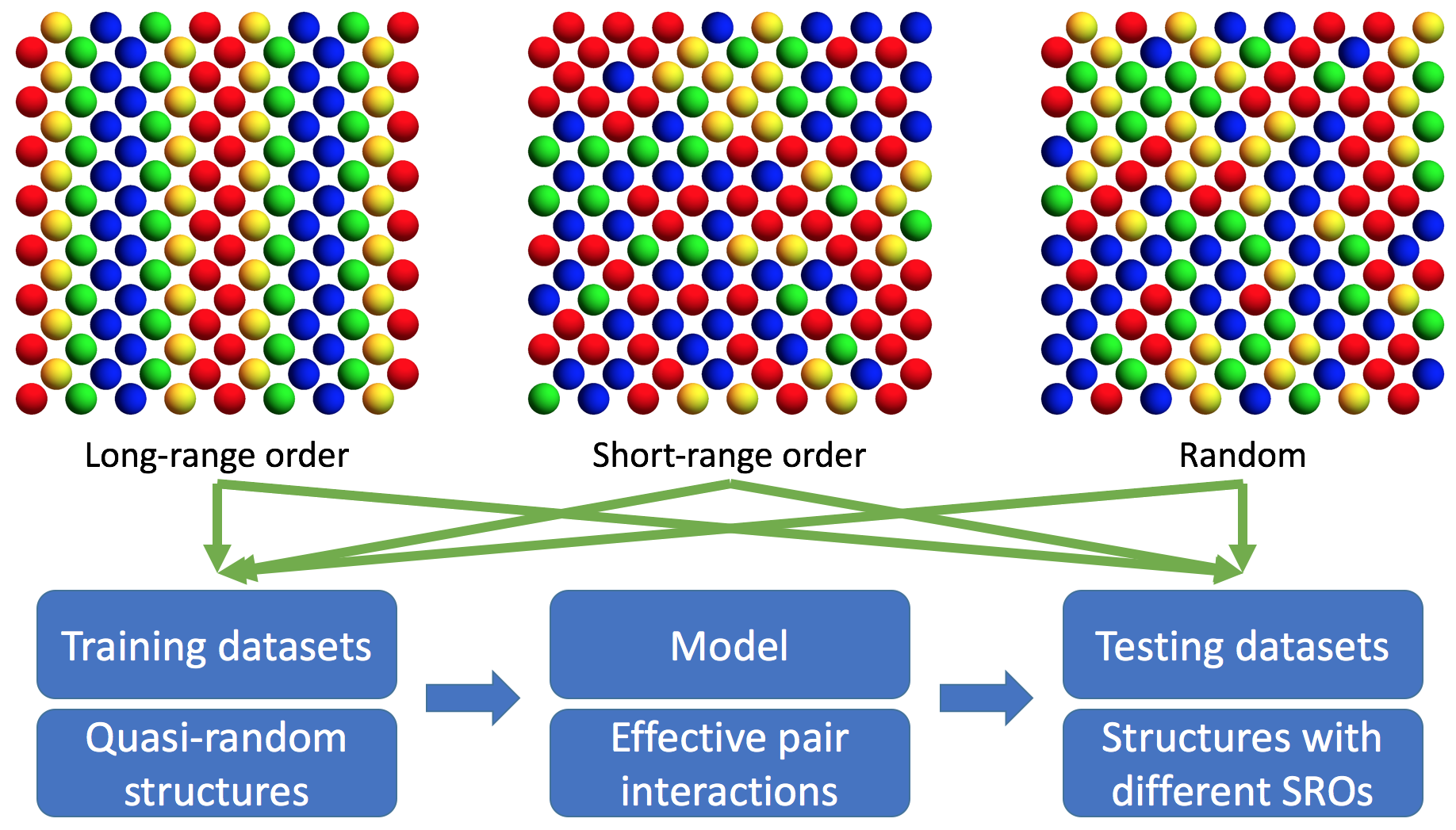}
\caption{ (Color online) A schematic to illustrate the main idea of our method. For a representative data, various degree of order should be included to the configurations in the data set.} \label{fig:schematic}
\end{figure}

To fulfill these requirements,  we first need a set of parameters to characterize the phase space and measure the ``distance" between different configurations. The short-range order (SRO) parameters are exactly such quantity. The SRO parameters can be measured in experiment via X-ray diffraction and are widely formulated in alloy theory. With the help of SRO parameters, the requirement of representative data is converted into spreading the data ``homogeneously" over the space spanned by the SRO parameters. Moreover, corresponding to each SRO parameter there is an effective pair interaction, in terms of which it is natural to construct the effective Hamiltonian. 
Such a method satisfies all the above three requirements, and is applied to study the prototypical NbMoTaW refractory HEA in this work. The schematic of our method is shown in Fig. \ref{fig:schematic}. The training data set is a range of quasi-random structures generated with different sizes of the supercell, which is a simple yet efficient technique to obtain training structures with various degrees of order and disorder. An effective pair interaction model is adopted to describe the configurational energies, with up to 8th nearest-neighbor shells included. The model is trained with linear regression on 1400 DFT data and tested with 200 structures of different SROs generated from simulated annealing, which is referred to as SRO data. On the other hand, the neural network methods are also employed for comparison.

\section{Results}

\subsection{Training and testing data}
\begin{figure}[h!]
\centering
   \includegraphics[width=0.9\textwidth]{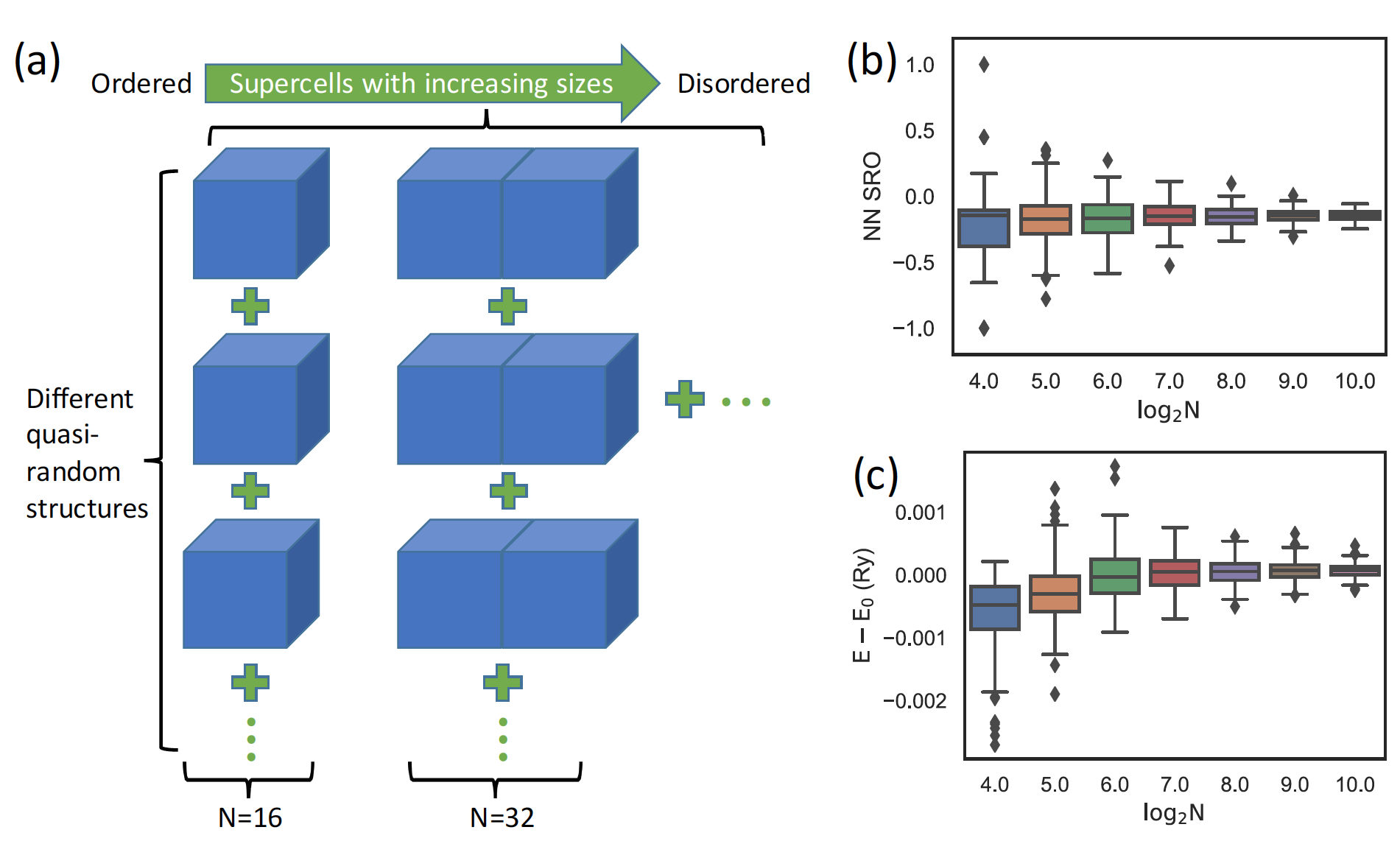}
\caption{ (Color online) (a) A schematic of to illustrate our method to obtain the training data set. (b) Box plot of the nearest-neighbor SRO parameters in the training data, for different supercell size N. (b) Box plot of the energies of the training data, for different supercell size N. } \label{fig:train}
\end{figure}

The energy of a condensed matter system can be split into two parts: one depends on the configuration (arrangement of atoms in perfect lattice), and the other depends on the positions of the atoms. The position part includes atomic relaxations and lattice vibrations, and are generally very complicated. The configuration part, on the other hand, can be described relatively easily with a lattice model, and produces the key differences between HEAs and traditional alloys, therefore will be the focus of our study. For simplicity, the atoms are assumed to be on the perfect cubic sites, which generally affects the low temperature ground state, but has less effect at other temperatures \cite{Korman_npj}. To characterize the different chemical configurations, the Warren-Cowley short-range order parameters \cite{PhysRev.77.669, OWEN2016155} is employed, which is defined as
\begin{equation}
\alpha_m^{AB} = 1 - \frac{P^{A|B}_m}{c_A}, \label{SRO}
\end{equation}
where $m$ represents the coordination shell, $c_A$ is the concentration of element $A$, and $P^{A|B}_m$ is the probability of finding element $A$ at the $m$-th neighbor shell of element $B$. For an  $n$-multicomponent alloy, a total of $n (n-1)/2$ nearest-neighbor SRO parameters exist at each shell. A negative value of $\alpha_m^{AB}$ indicates the preference of forming $AB$ bonds at the $m$-th shell, while positive $\alpha_m^{AB}$ suggests the opposite, and $\alpha_m^{AB}=0$ for each $m$ corresponds to a completely random system. 


Other than short-range order, the training data set also benefit from incorporating different long-range order. A simple yet efficient strategy to achieve both is to combine the DFT data calculated with different sizes of the supercell. Because of the periodic boundary conditions, the samples obtained with small supercells already include different long-range order, while the configurations from larger supercells contain various degree of short-range order, as illustrated in Fig. \ref{fig:train}(a). From Fig. \ref{fig:train}(b) and  \ref{fig:train}(c)  it is easy to see that mixing these data produces an ensemble of a wide range of SRO and energy values. Of course, instead of quasirandom structures, one may further improve the quality of the data by generating configurations of different SROs with simulated annealing. This is the approach we employ to obtain the testing data. Nevertheless, as will be shown in the results, the simple mixing of quasirandom structures already produces high quality training data.

\begin{figure}[!ht]    
    \centering
    \subfigure[]{\includegraphics[width=0.3\textwidth]{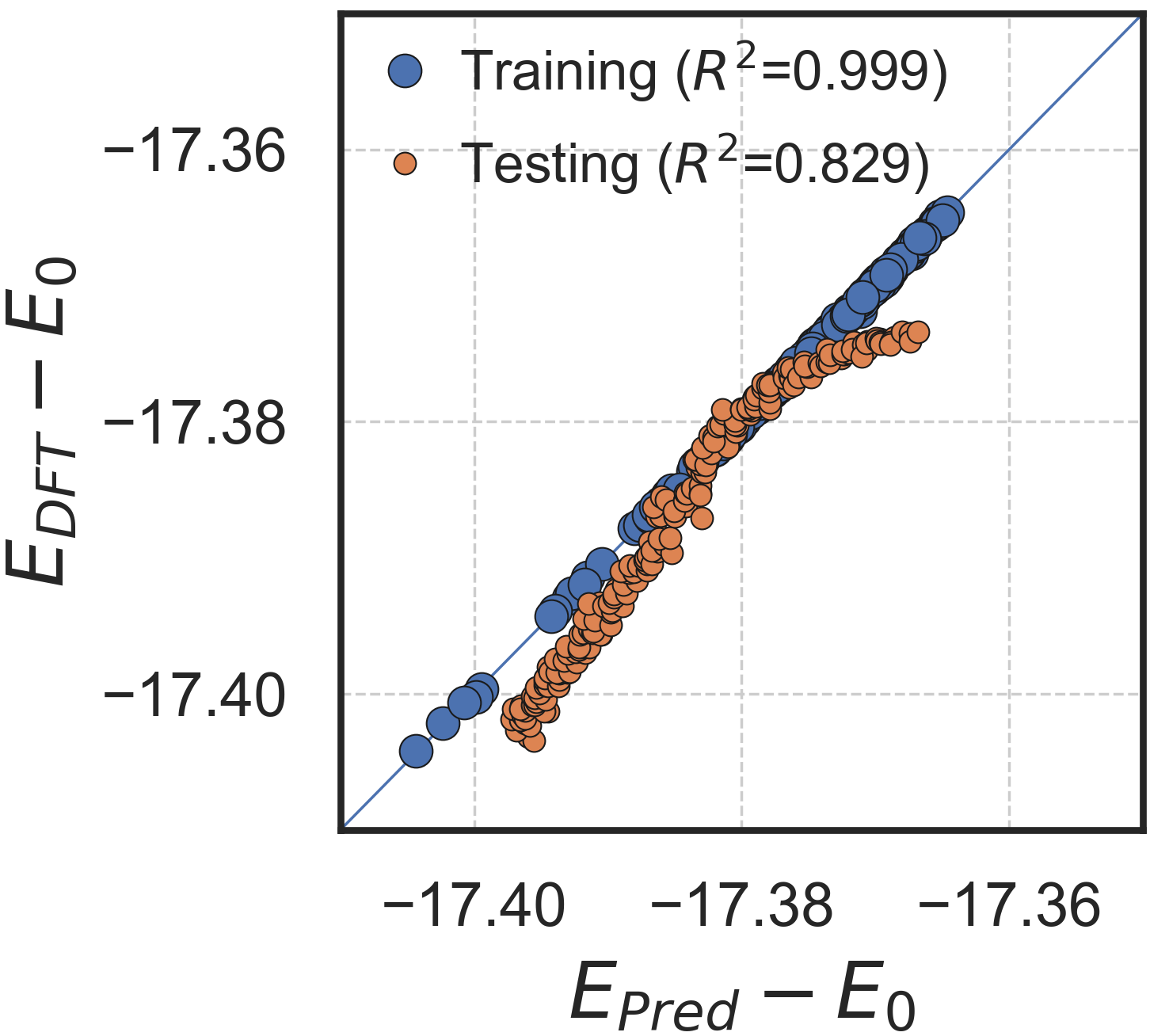}}
    \subfigure[]{\includegraphics[width=0.3\textwidth]{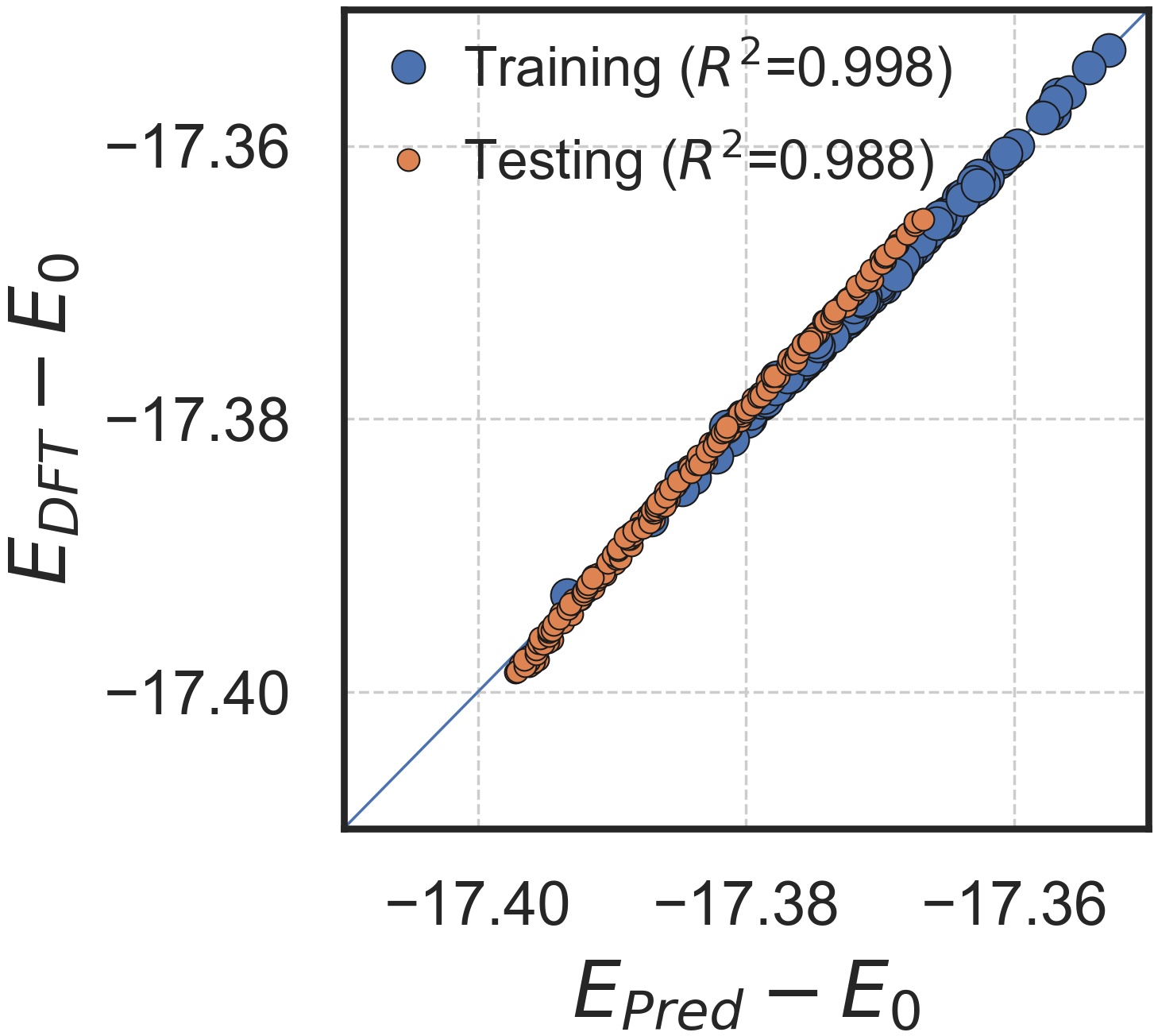}}
    \subfigure[]{\includegraphics[width=0.3\textwidth]{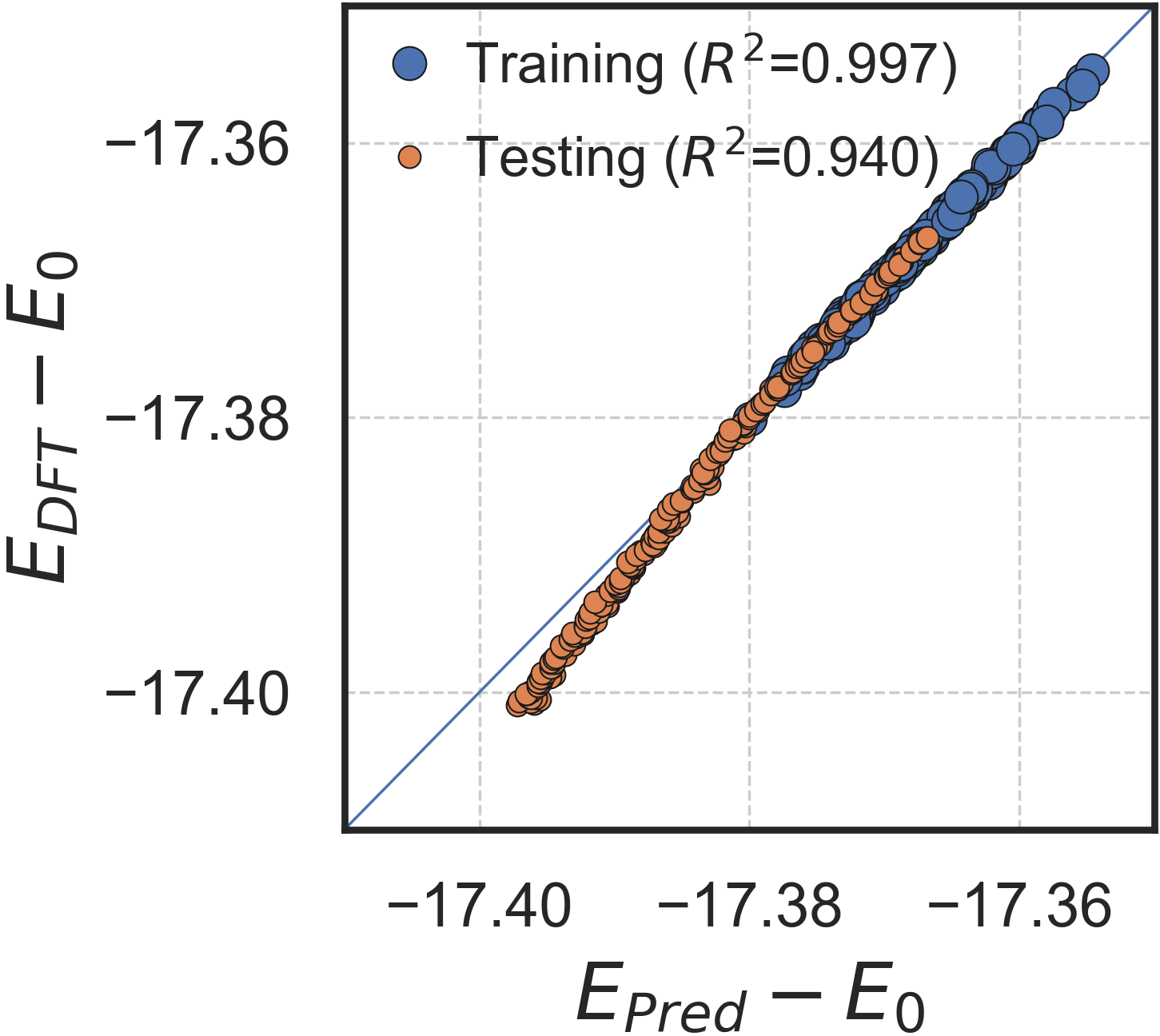}}
    \subfigure[]{\includegraphics[width=0.3\textwidth]{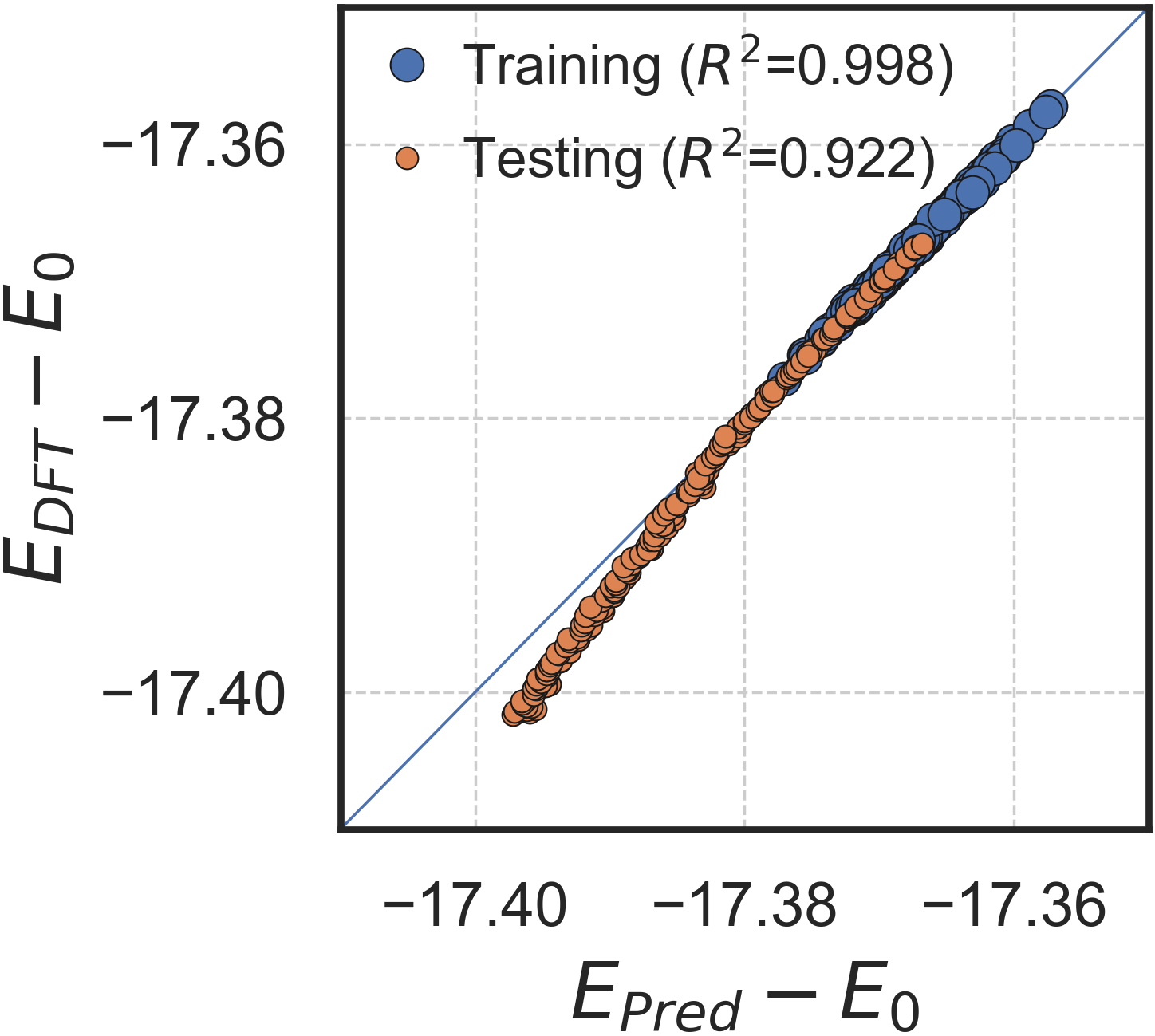}} 
        \subfigure[]{\includegraphics[width=0.3\textwidth]{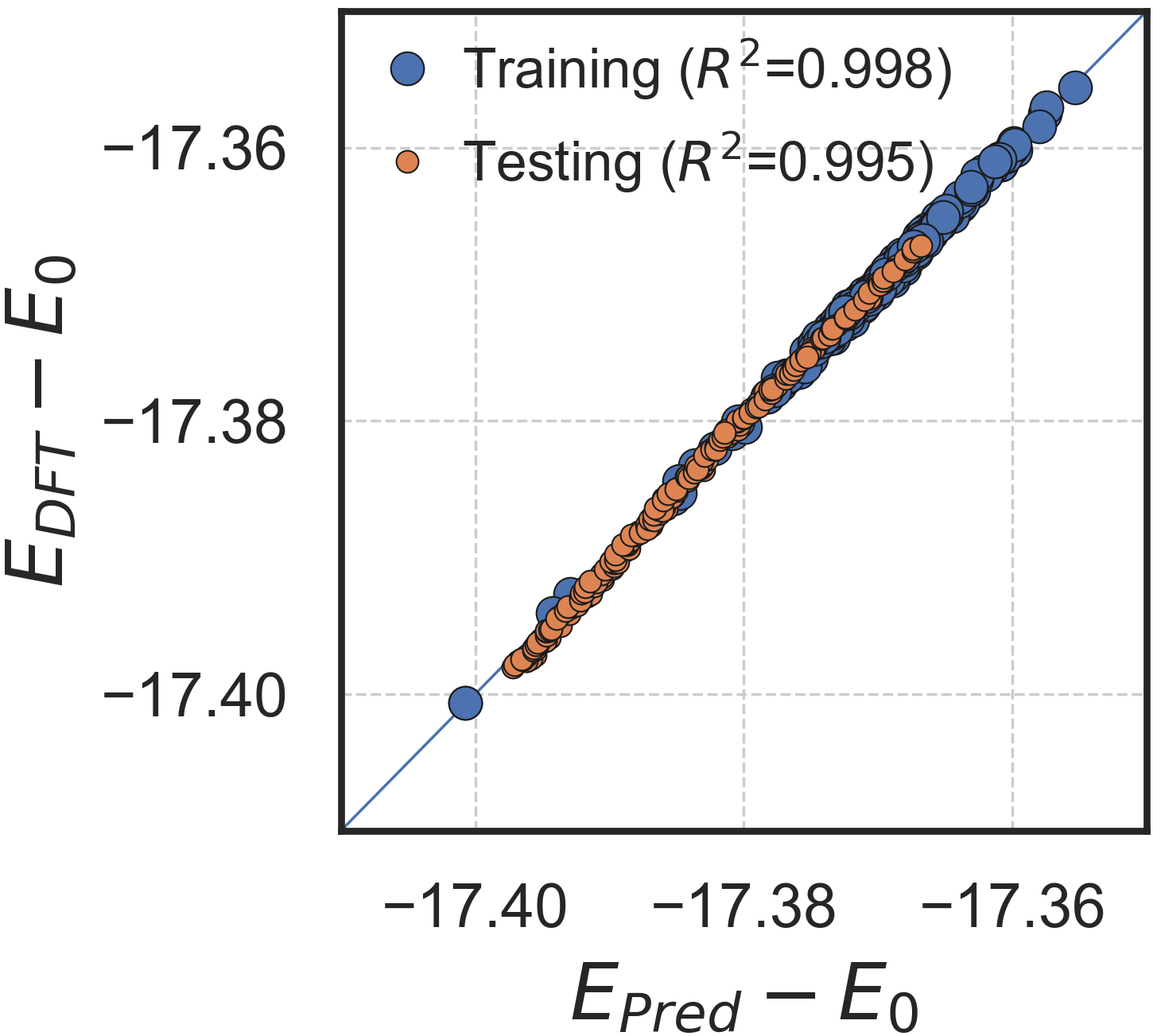}} 
    \caption{Machine learning prediction results given different sources of training data set (a) 200 training data from 16-atom supercell, (b) 200 training data from 32-atom supercell, (c) 200 traing data from 64-atom supercell, (d) 200 training data from 128-atom supercell and (e) 200 ensemble training data from 16, 32, 64 and 128-atom supercell (randomly draw 50 data from each supercell size). $E_{DFT}$ and $E_{pred}$ are referred to as the DFT-based prediction and ML-based prediction respectively and $E_0$ is a constant.} 
     \label{fig:f3}
\end{figure}

\begin{figure}[!ht]    
    \centering
    {\includegraphics[width=0.45\textwidth]{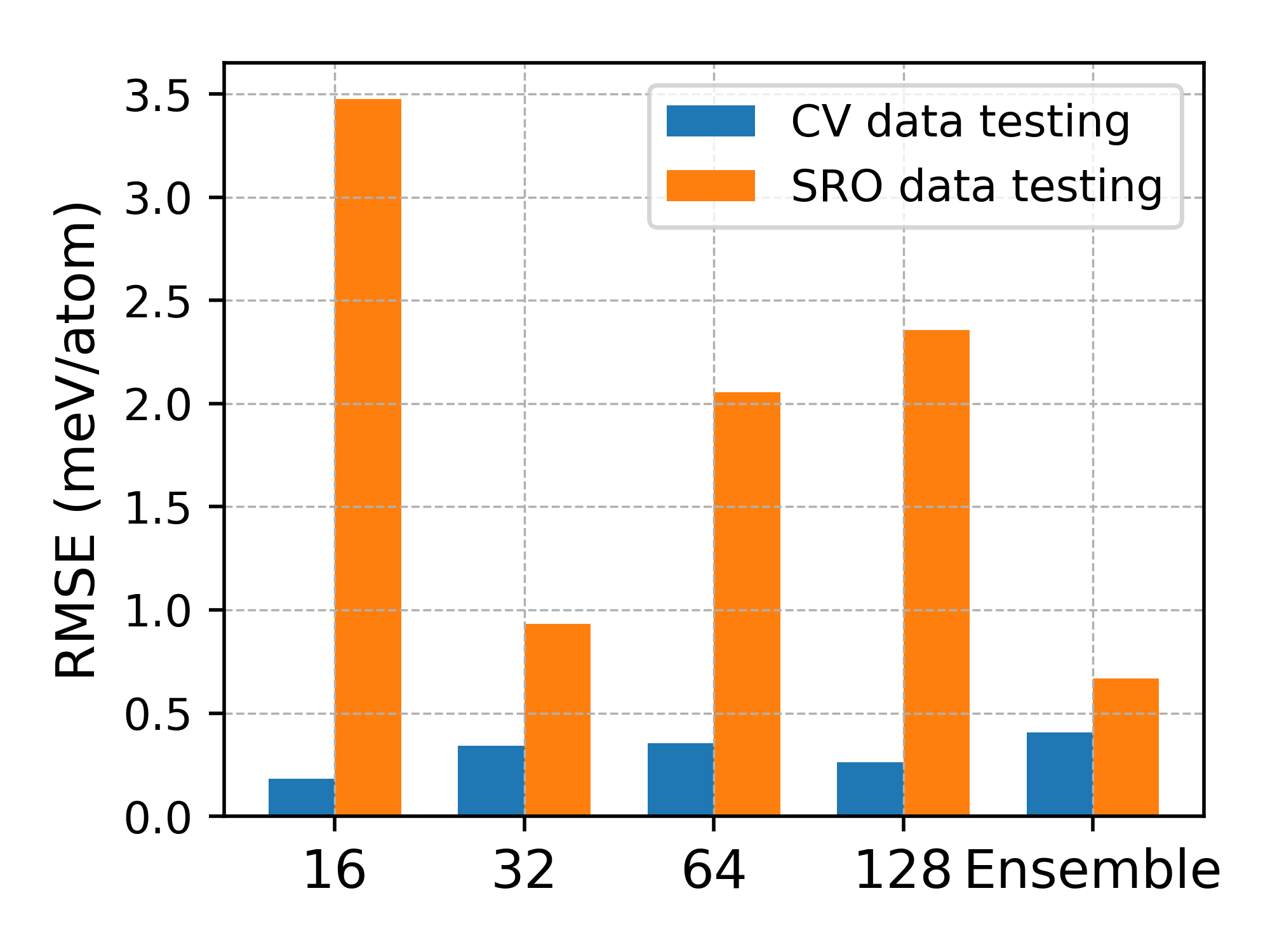}} \quad
    \caption{Effect of testing data on prediction accuracy (RMSE). Blue bars: cross-validation testing results given 200 data drawn from 16, 32, 64, 128-atom supercell and ensemble data set. Orange bars: short-range order (SRO) testing data results using well-training model with 200 data drawn from 16, 32, 64, 128-atom supercell and ensemble data set (mixing of the data)}\label{fig:f4}
\end{figure}

The benefit of mixing the training data can be easily seen in the test as shown in Fig. \ref{fig:f3}, where the training and testing results of 16, 32, 64, 128-atom supercells are compared with the that of the mixing data. The number of training data is chosen as the same (200) to exclude the effect of sample size, and the model parameters are also set as the same (coordination shell $m$ = 6). For the 16-atom supercell, the training score $R^2 = 0.999$ show that the trained model is very accurate for a system described by a 16-atom supercell. However, the $R^2$ testing score obtained using the 1024-atom SRO data is only 0.829, which demonstrates the incompetence of the 16-atom data to generalize over the whole configuration space. This result also underscores the danger of representing a random system with a small supercell, which is a common practice due to its efficiency, but can lead to loss of physical information at the thermodynamic limit. The results of 32, 64, and 128-atom are better than that of 16-atom, reflecting that more physics is captured by these training data. Nevertheless, all of them are outperformed by the mixing data set, which is an ensemble drawn from the 16, 32, 64, and 128-atom data. The mixing data set produces a testing score of 0.995 and root mean square error (RMSE) of 0.6 meV. Moreover, from Fig. \ref{fig:f4}, it can be seen that the cross validation testings constantly underestimate the RMSE. If no other independent testing, this would lead to the wrong conclusion that a ``good” model has been successfully trained. To avoid such a pitfall, the correct practice should be using an independent testing data set, which is the SRO data in our case. From the RMSE of the SRO data, again we see that the mixing data set beats the others in terms of the accuracy and reliability. In addition, for the mixing data set, the errors from cross validation and SRO data testing have similar magnitude, which further verified its representativeness.

\subsection{Model}
As mentioned in the introduction, the traditional cluster expansion method is generally difficult when applied to multicomponent alloys because the number of parameters scales as $n^{b}$ for a $b$-body cluster of $n$ elements. Therefore, unless the high-order interactions in the system are important, it would be preferable to use an Ising-like model with only effective pair interactions (EPIs). In terms of the EPIs, the effective Hamiltonian at lattice site $i$ can be written as \cite{PhysRevB.55.856}
\begin{equation}
H(i) = \sum_{j \neq i} V_{m}^{A(i)B(j)} c_j + V_0^{A(i)}, \label{Hamiltonian}
\end{equation}
where $A(i)$ represents element A at site $i$, $V_{m}^{A,B}$ is the interatomic pair potential between element A and B. $m$ is the coordination shell number signifying the separation between $i$ and $j$, $c_j$ is the occupation number, and $V_0$ is the concentration dependent part that can be discarded for a given composition. The total energy of the system can then be obtained by summing up the Hamiltonian over all atomic sites, which is given by:
\begin{equation}
E =N \sum_{A, B, m} V_{m}^{A,B} \sigma_{m}^{A,B}, \label{energy}
\end{equation}
where $N$ is the number of atoms in the system, and $\sigma_{m}^{A,B}$ is the percentage of $AB$ bonds in the $m$-th shell. For an $n$-component system, the number of different bonds in a single shell is $n(n+1)/2$. However, for a fixed chemical composition, there are also $n$ constranints from the concentration of each element, so the total number of independent parameters is $n(n-1)/2$. For instance, for the  NbMoTaW refractory HEA, there are 10 different bonds for each coordination shell., i.e., NbNb, NbMo, NbTa, NbW, MoMo, MoTa, MoW, TaTa, TaW, WW, but only 6 of the corresponding SRO parameters are independent. Therefore, Eq. \eqref{energy} can be written as
\begin{equation}
E =N \sum_{A>B, m} V_{m}^{A,B} P_{m}^{A|B}, \label{energy}
\end{equation}
Where $P_{m}^{A|B}$ is closely related to the short-range order parameter, as defined in Eq. \eqref{SRO}.
The parameters in Eq. \eqref{energy} can be obviously determined with linear regression using $P_{m}^{A,B}$ as the features. This would correspond to a truncation of the ECIs in cluster expansion to retain only the pair interactions.

\begin{figure}[!ht]    
    \centering
    \subfigure[]{\includegraphics[width=0.3\textwidth]{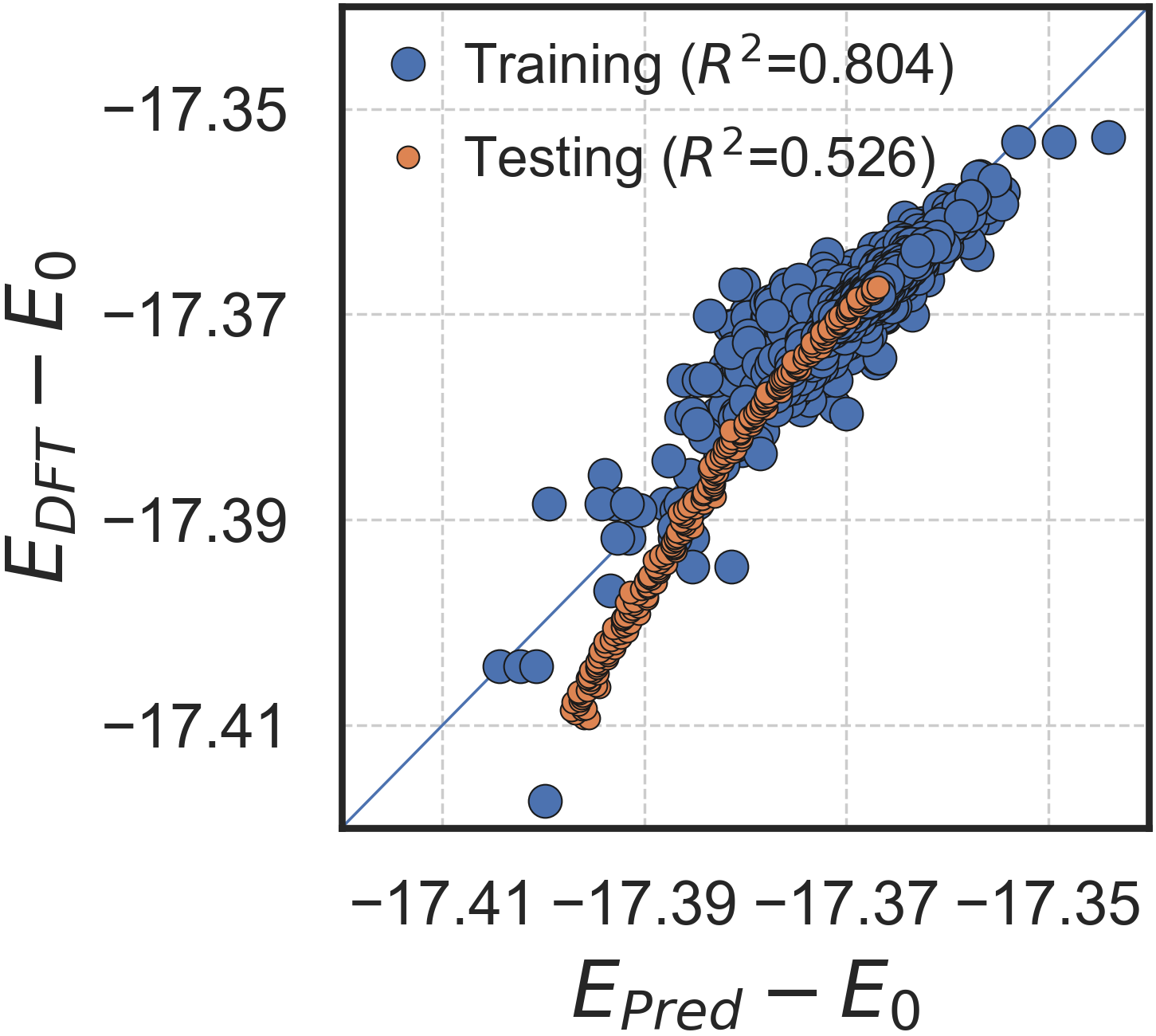}}
    \subfigure[]{\includegraphics[width=0.3\textwidth]{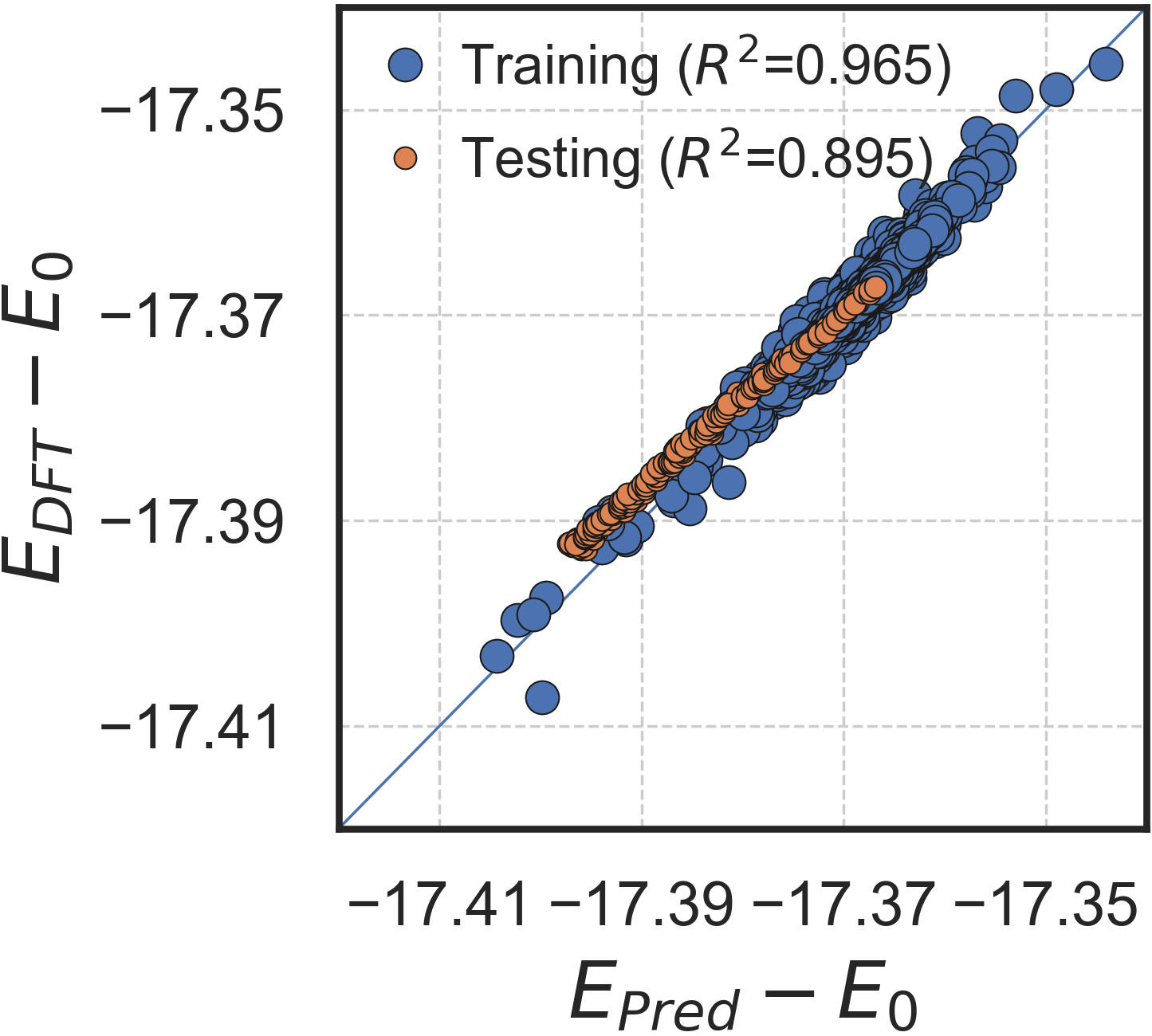}}
    \subfigure[]{\includegraphics[width=0.3\textwidth]{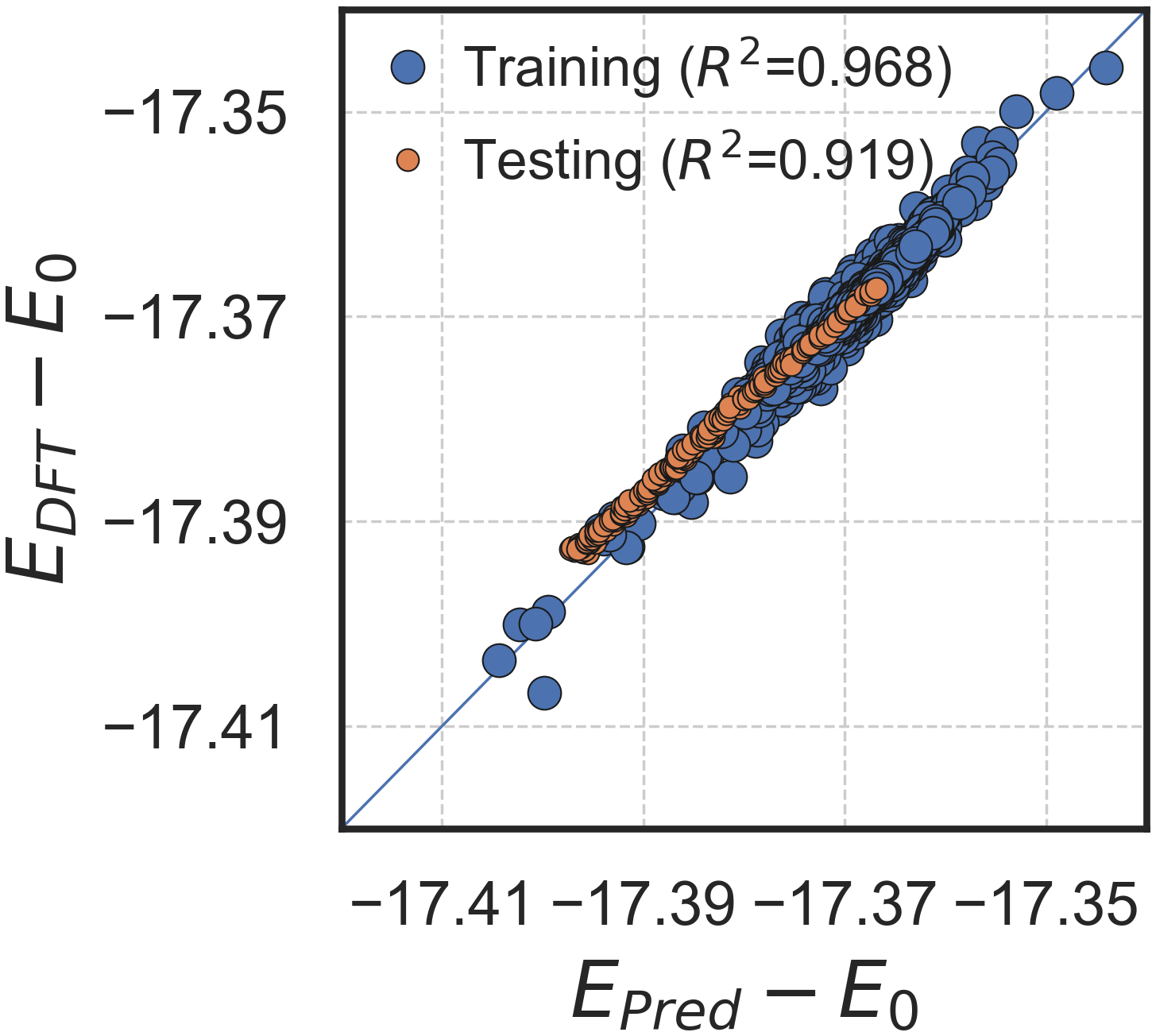}}
    \subfigure[]{\includegraphics[width=0.3\textwidth]{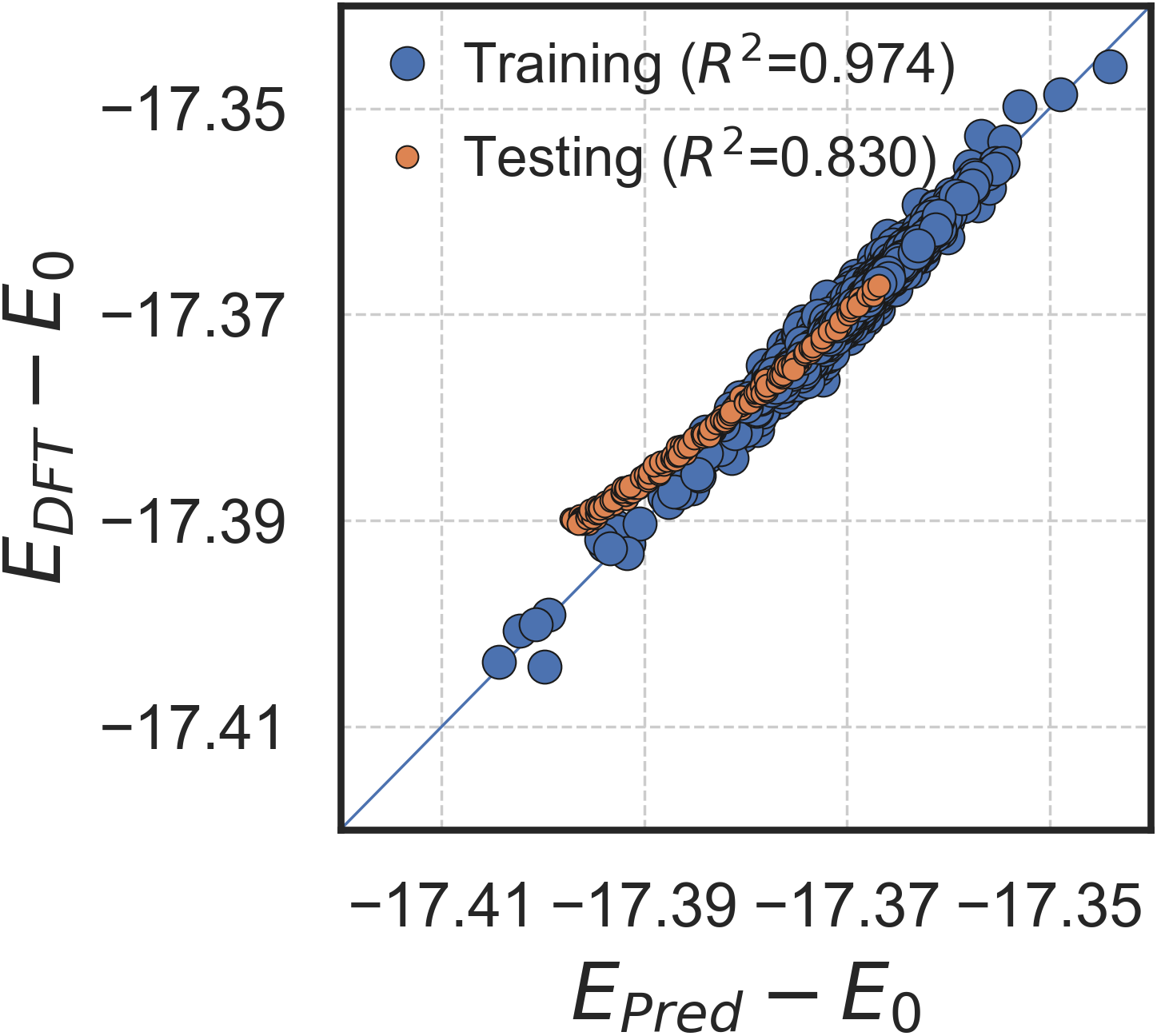}} 
\subfigure[]{\includegraphics[width=0.3\textwidth]{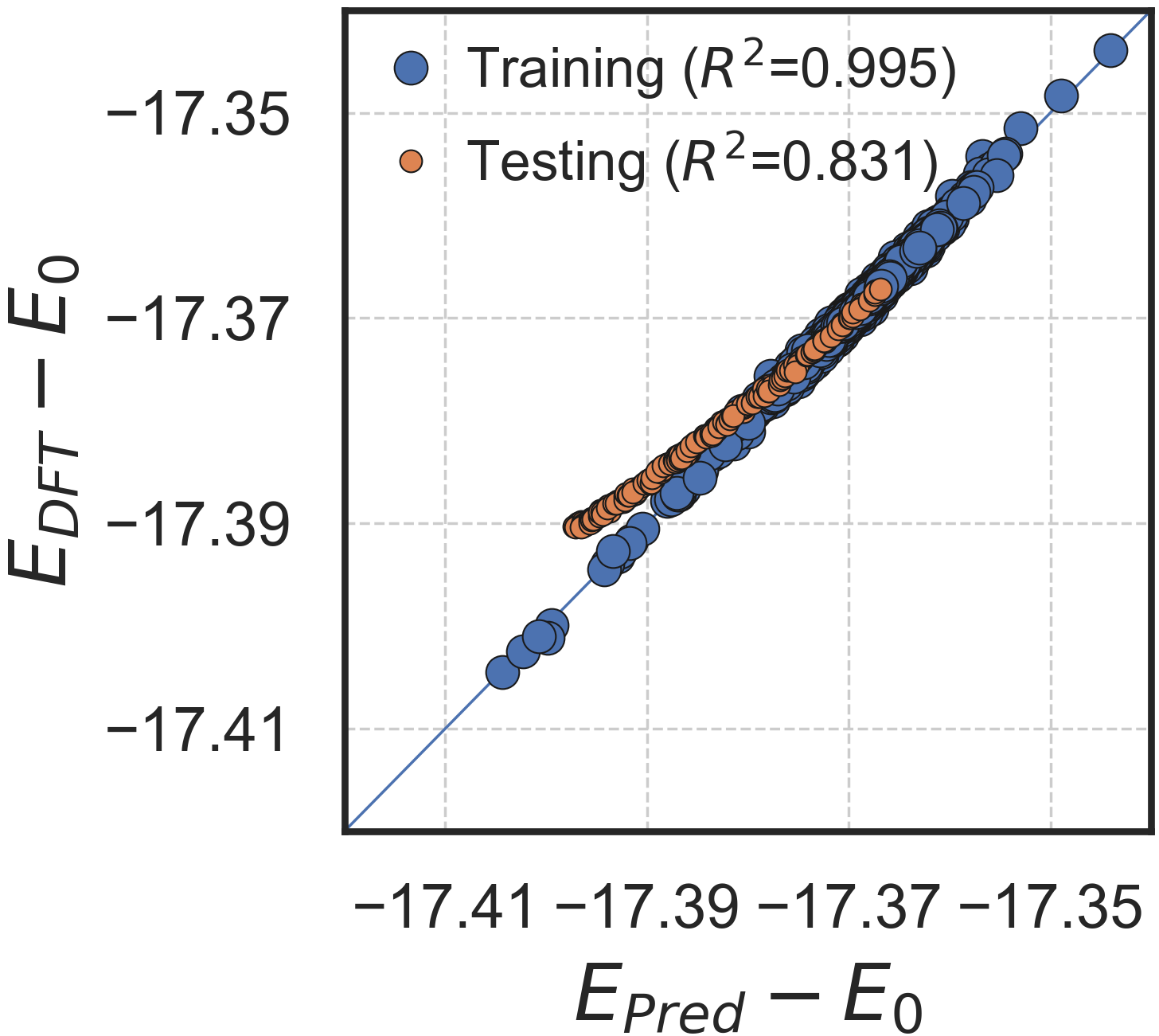}}
\subfigure[]{\includegraphics[width=0.3\textwidth]{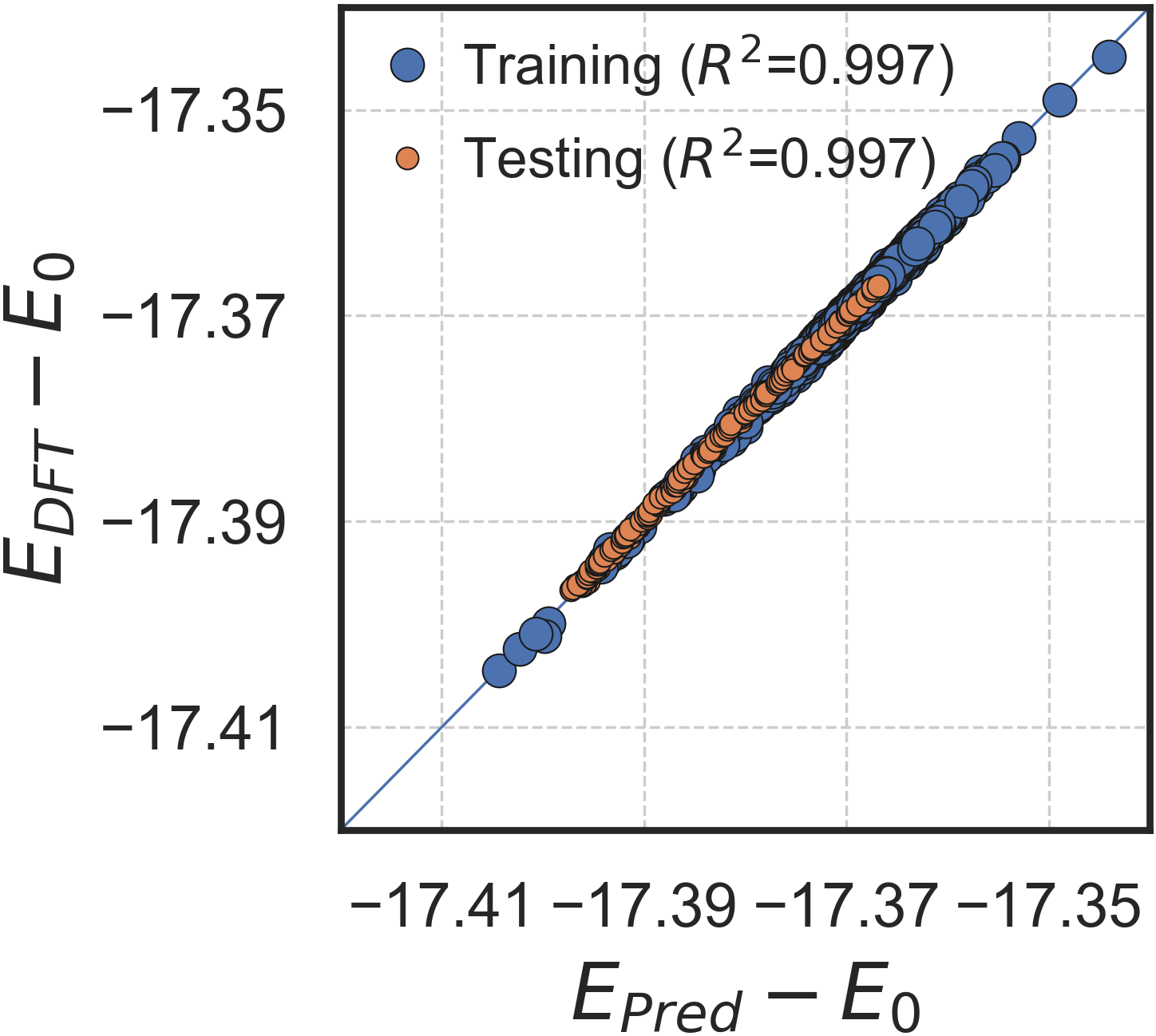}}
\subfigure[]{\includegraphics[width=0.3\textwidth]{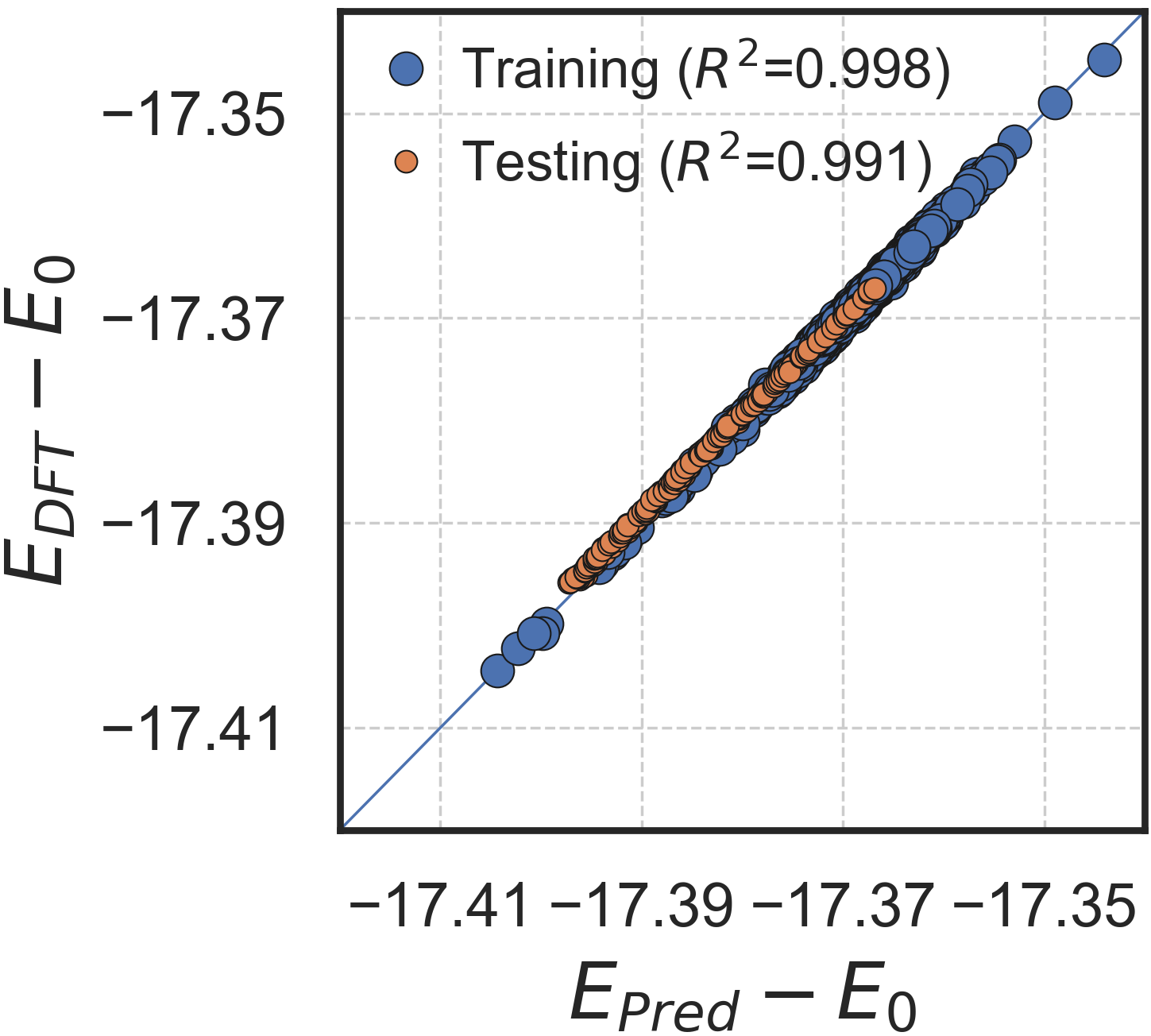}}
\subfigure[]{\includegraphics[width=0.3\textwidth]{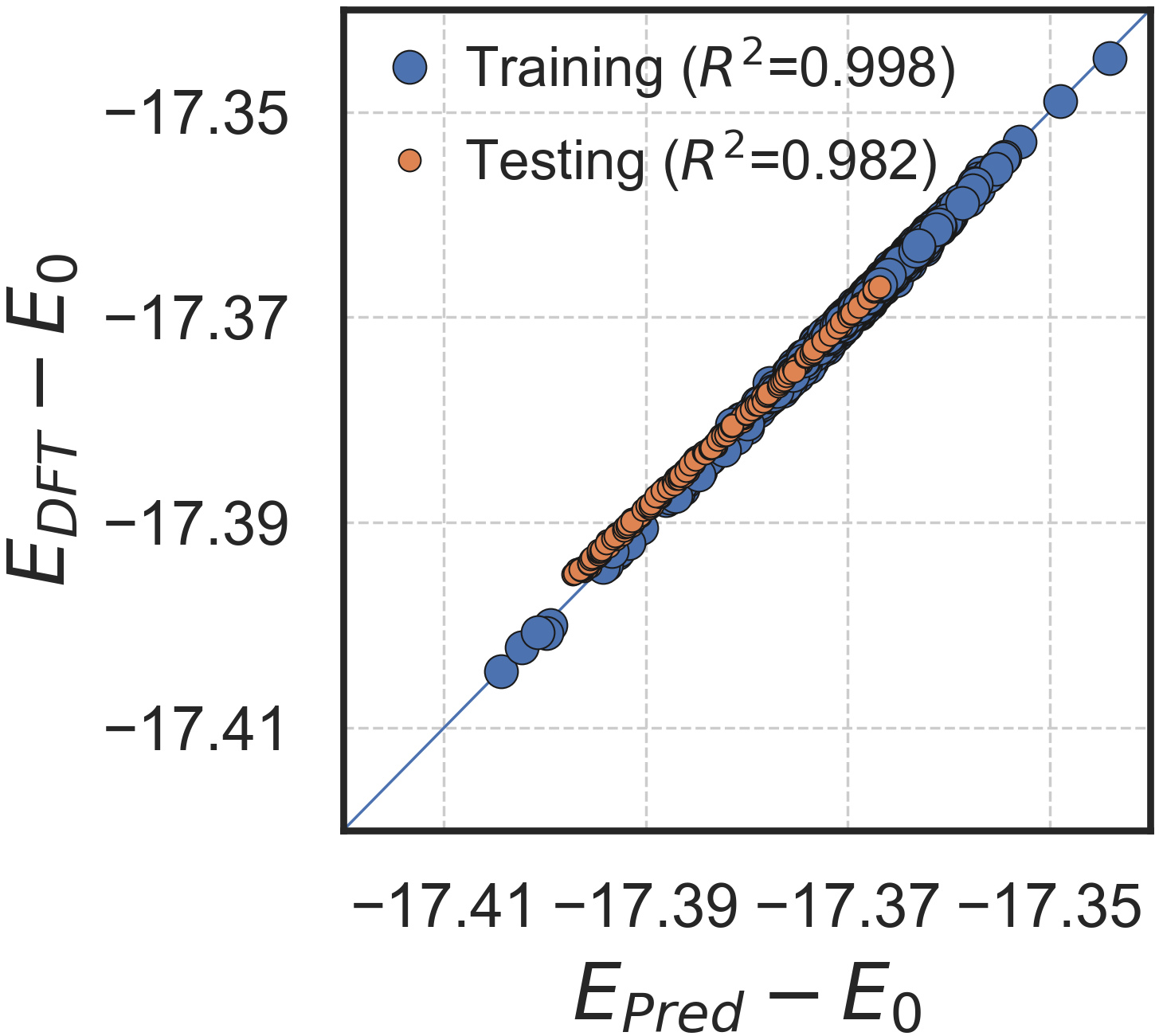}}
    \caption{Effect of the number of coordination shells on prediction performance. (a)-(h) corresponds to the 1-8 coordination shells respectively. The blue dots and orange dots represent the 1400 training data and 200 short-range order testing data}  \label{fig:f5}
\end{figure}

\begin{figure}[!ht]    
    \centering
    {\includegraphics[width=0.4\textwidth]{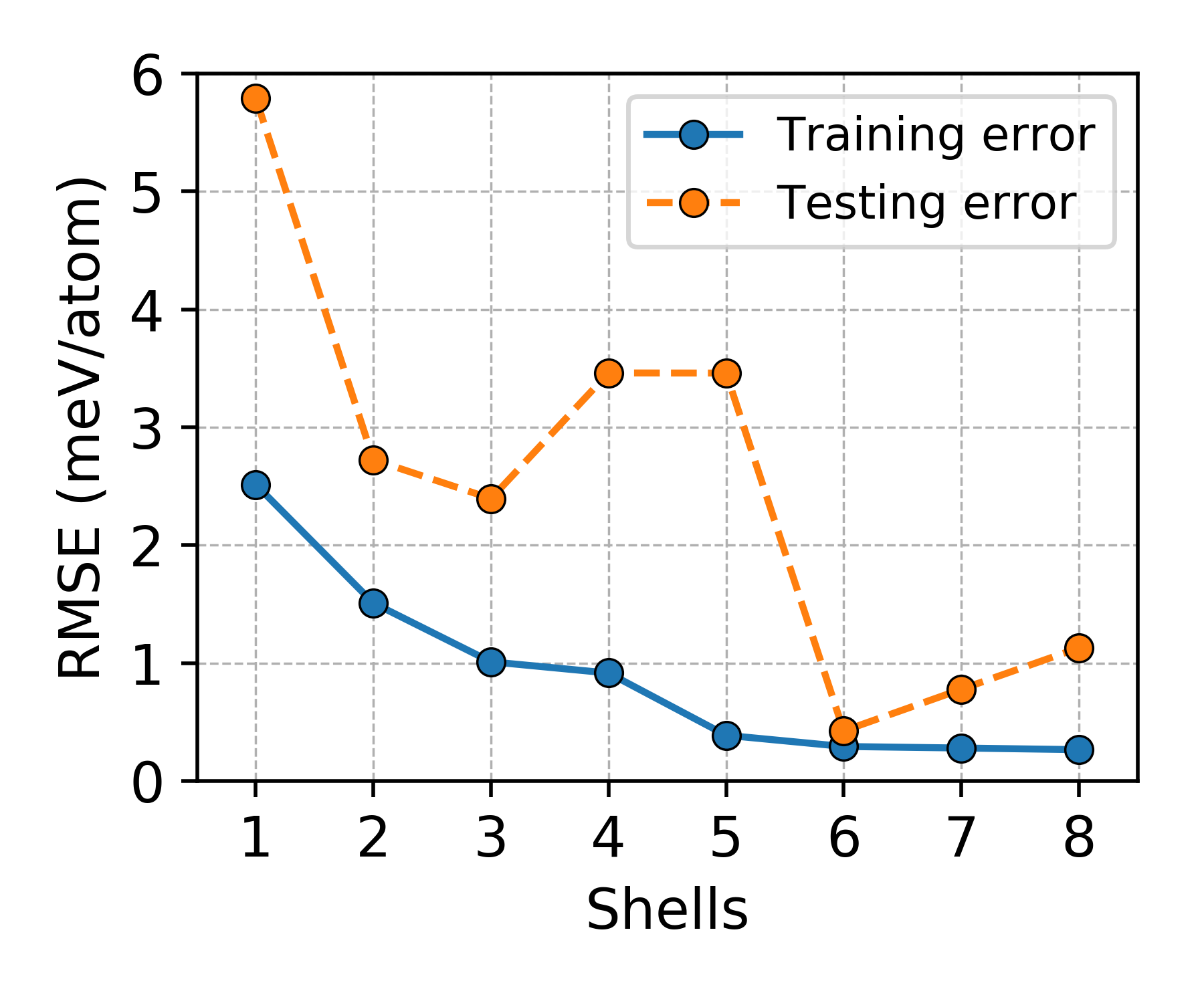}} \quad
    \caption{Effect of the number of coordination shells on training accuracy and short-range order (SRO) testing accuracy (RMSE)} \label{fig:f6}
\end{figure} 

In practice, the cutoff for the number of coordination shells, $m$, need to be examined and determined for each specific material. For the NbMoTaW HEA, the impact of $m$ on the accuracy of the trained models is shown in Fig. \ref{fig:f5} and Fig. \ref{fig:f6}. As can be seen, the nearest-neighbor model gives a rough approximation, with a testing score of 0.526 and RMSE of 5.79 meV. After including the next nearest-neighbor pair interactions, the accuracy of the model is substantially improved, with the testing score increased to 0.895 and RMSE reduced 2.72 meV. As the number of shells further increases, the training accuracy generally becomes better, but the testing accuracy reaches a maximum at the 6th shell, after which including more shells into the effective Hamiltonian would reduce the testing score, indicating that too many features in the model could lead to poor performance due to the bias-variance tradeoff. By including the first 6 shells of pair interactions, the model Hamiltonian demonstrates an accuracy of 0.425 meV when compared against the DFT results.

\begin{figure}[!ht]    
    \centering
    \subfigure[]{\includegraphics[width=0.45\textwidth]{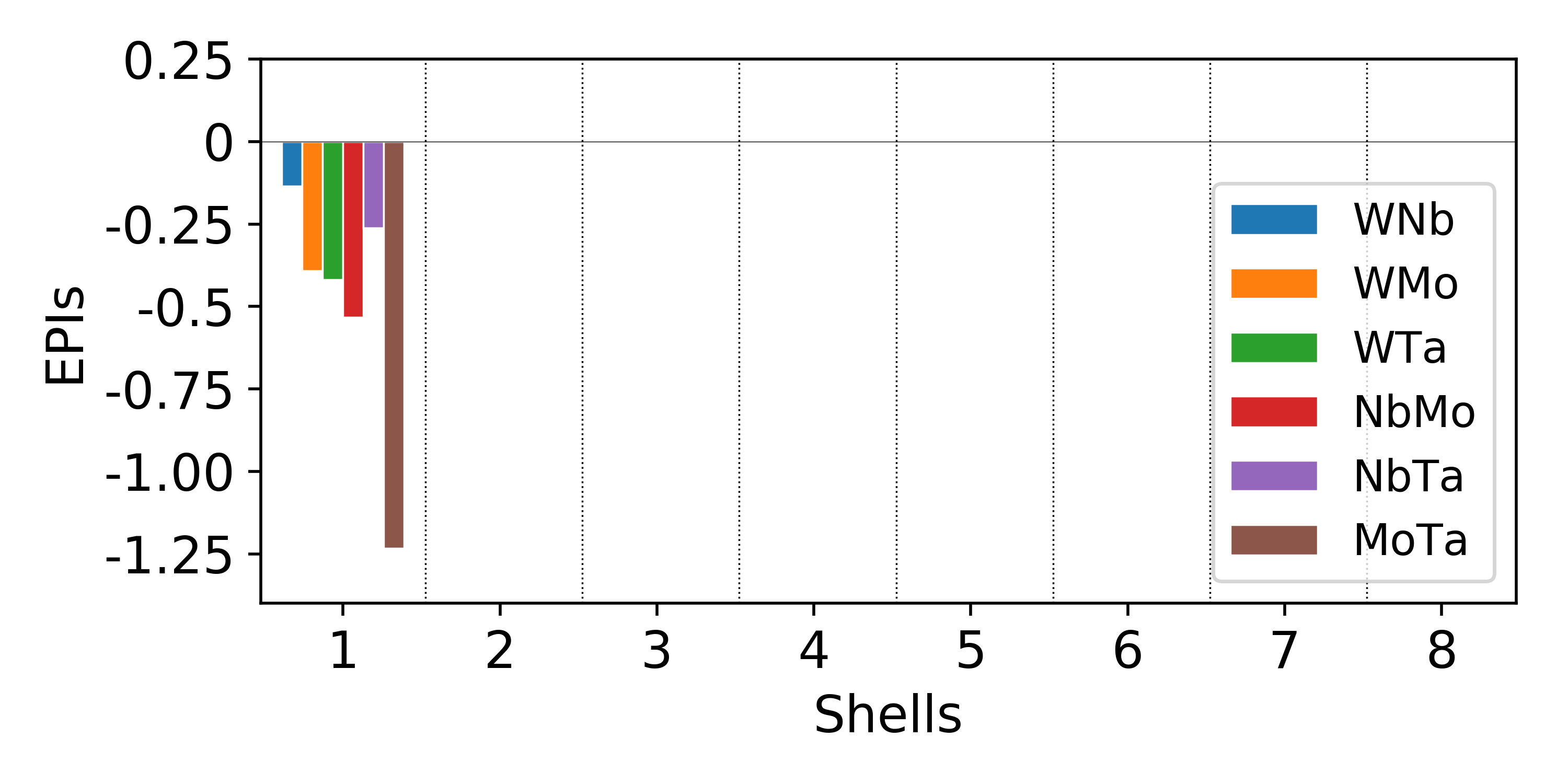}}
    \subfigure[]{\includegraphics[width=0.45\textwidth]{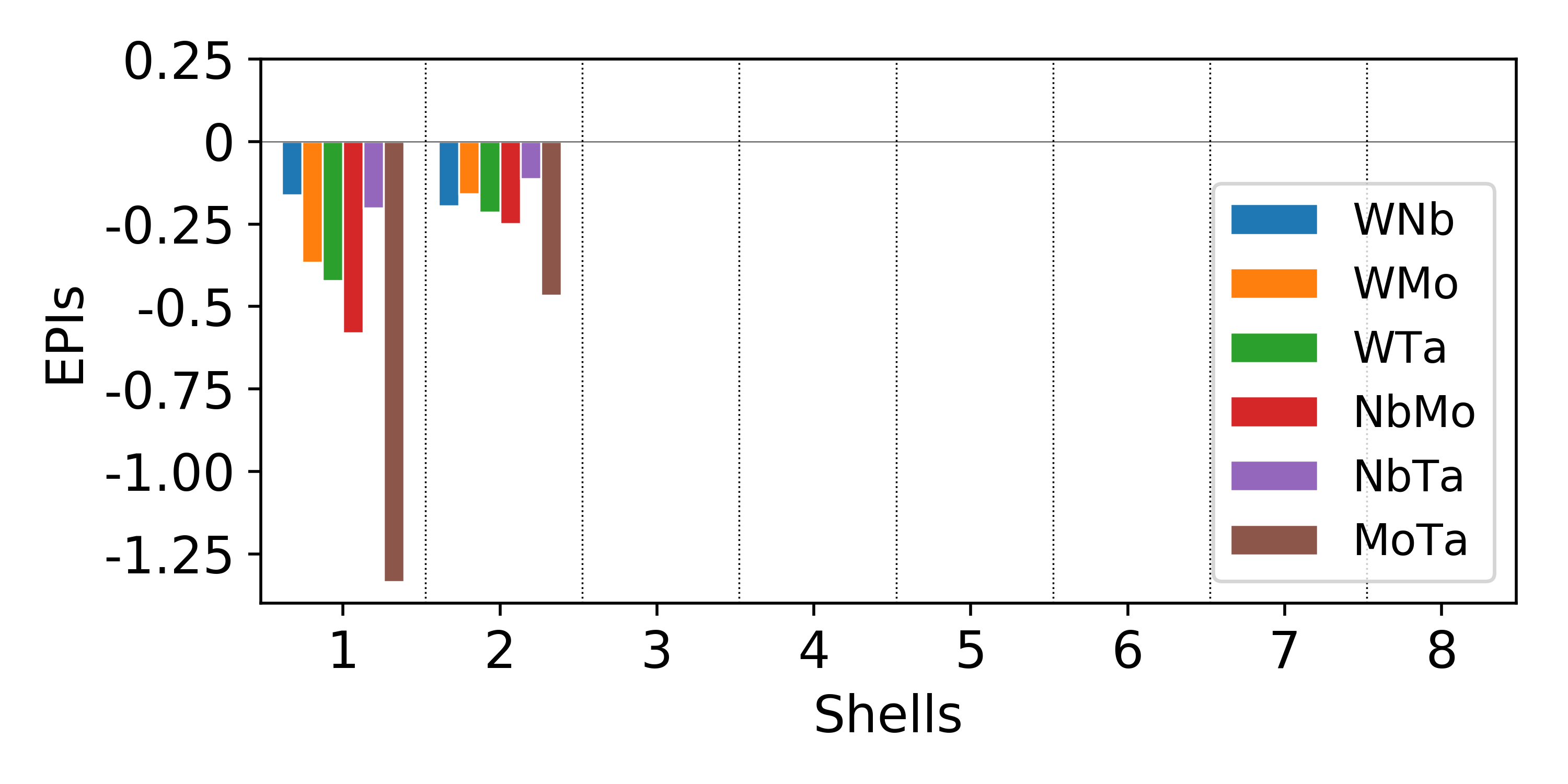}}
    \subfigure[]{\includegraphics[width=0.45\textwidth]{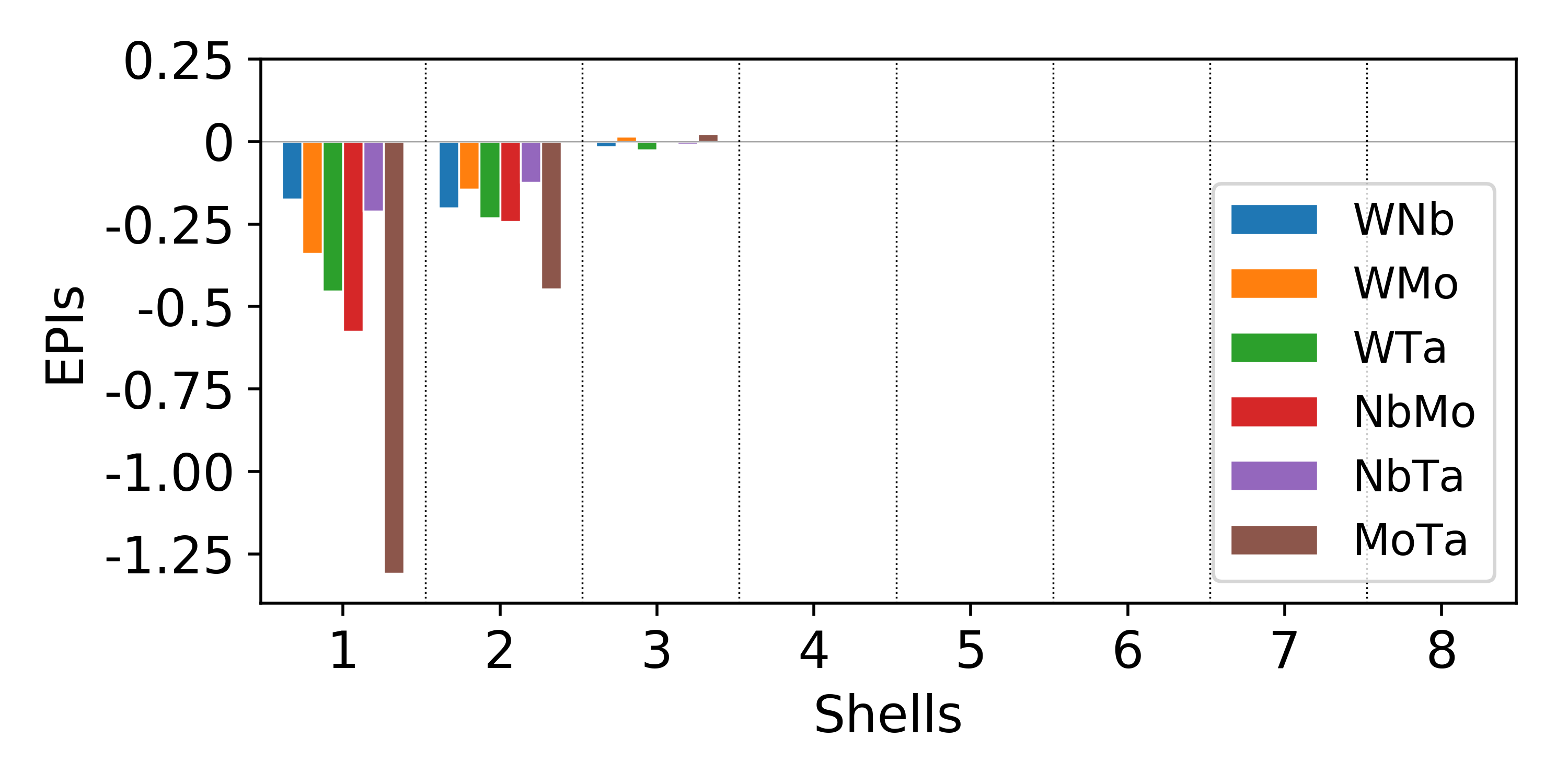}}
    \subfigure[]{\includegraphics[width=0.45\textwidth]{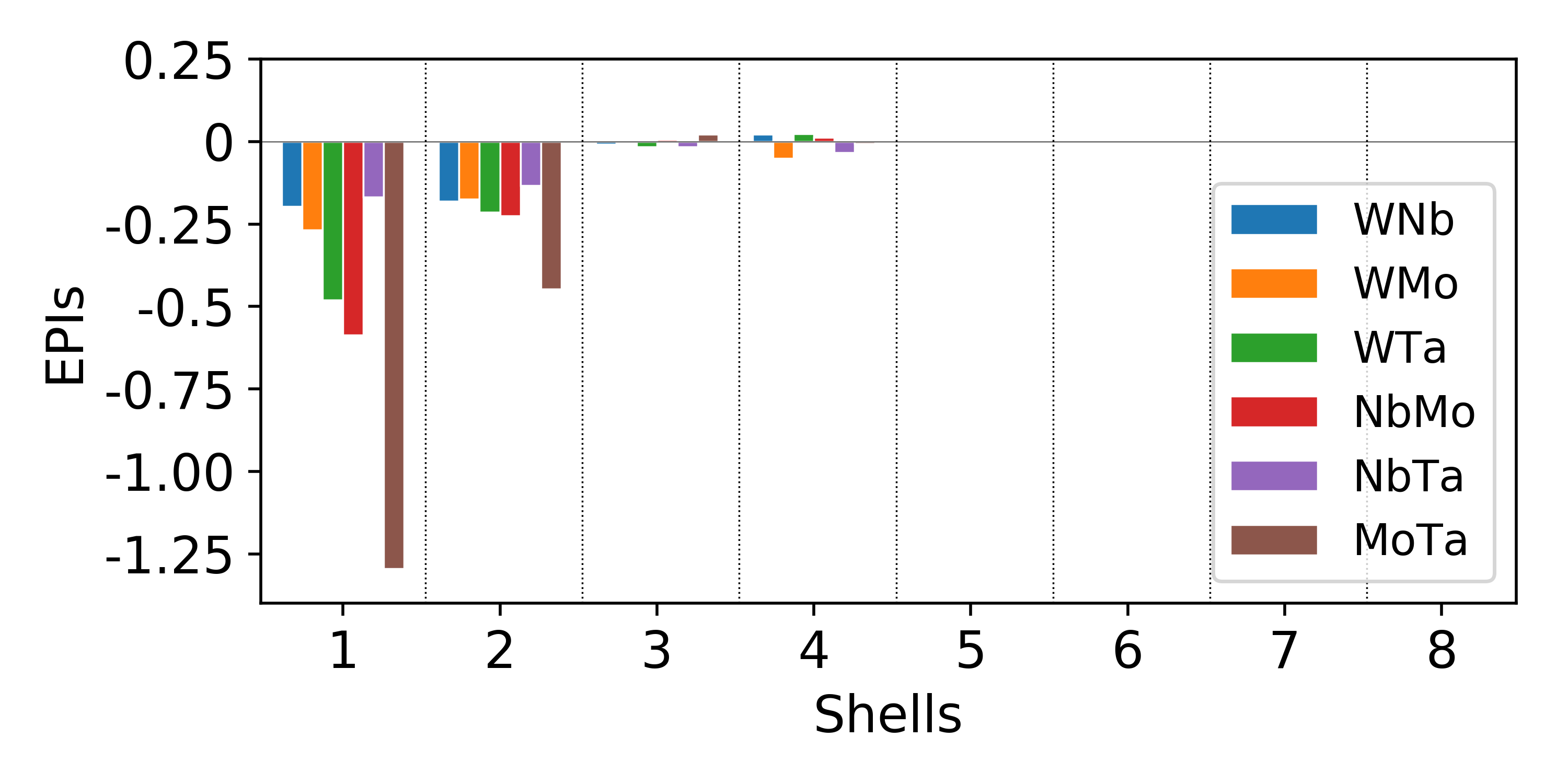}}
\subfigure[]{\includegraphics[width=0.45\textwidth]{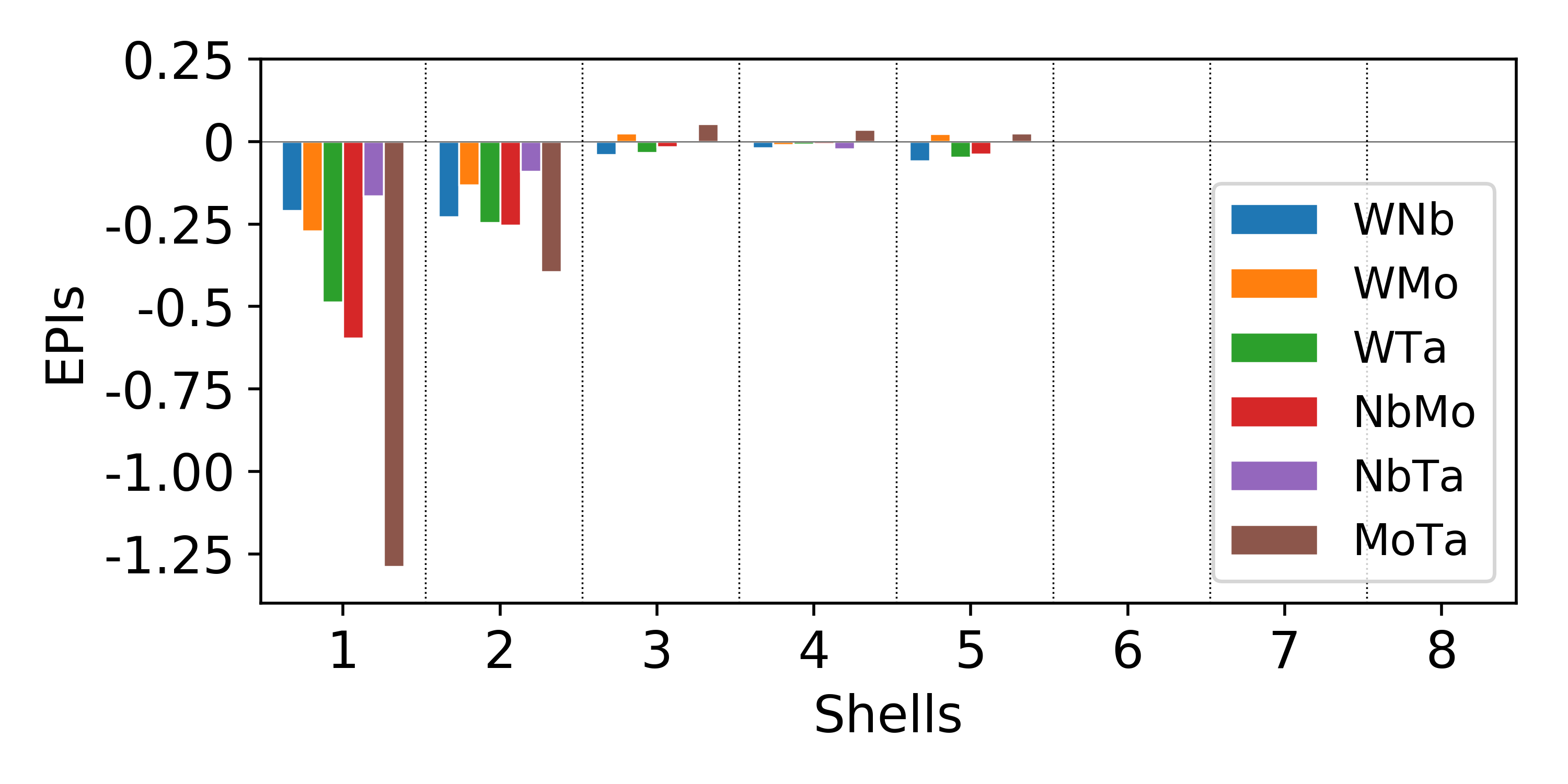}}
\subfigure[]{\includegraphics[width=0.45\textwidth]{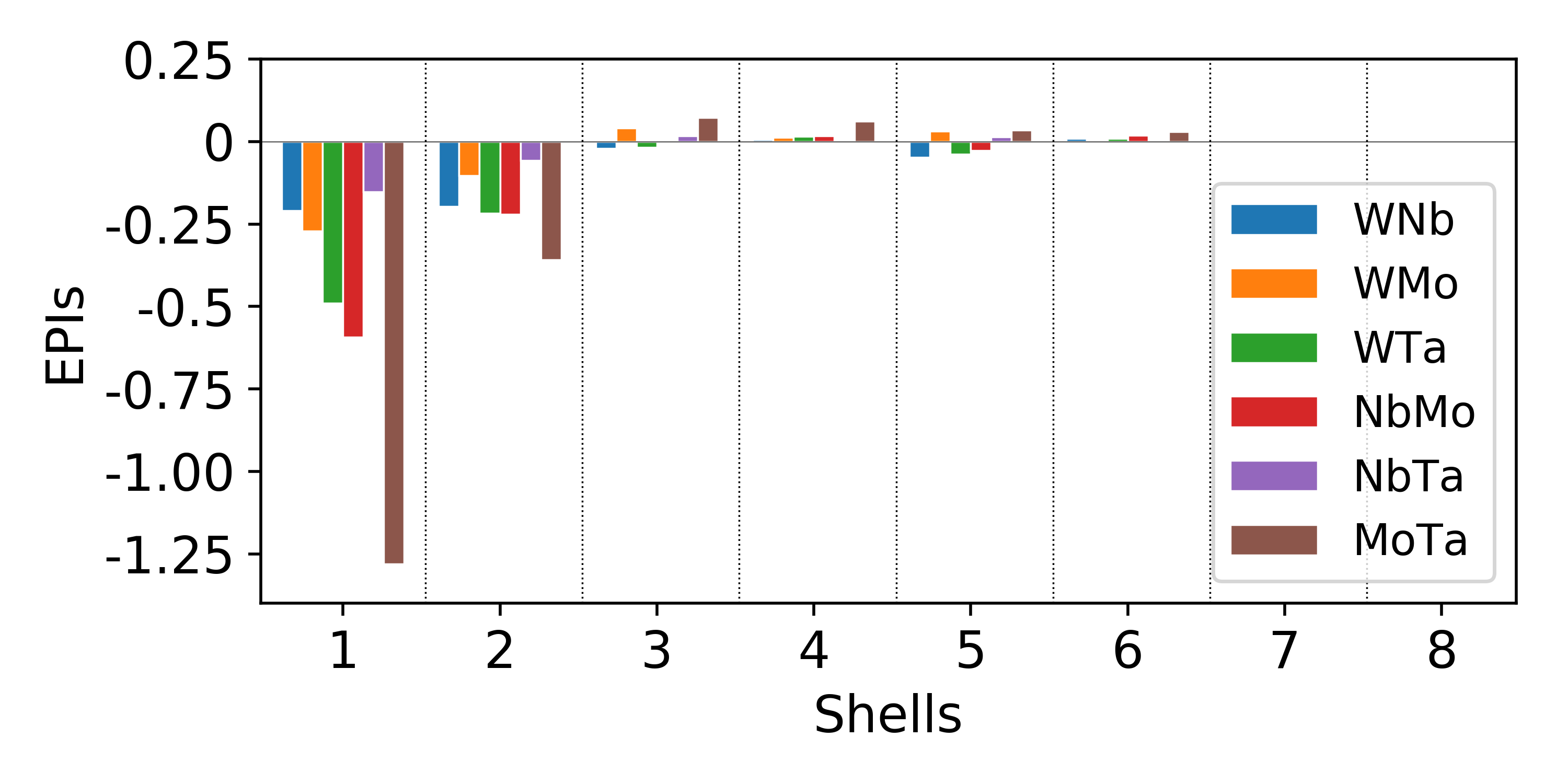}}
\subfigure[]{\includegraphics[width=0.45\textwidth]{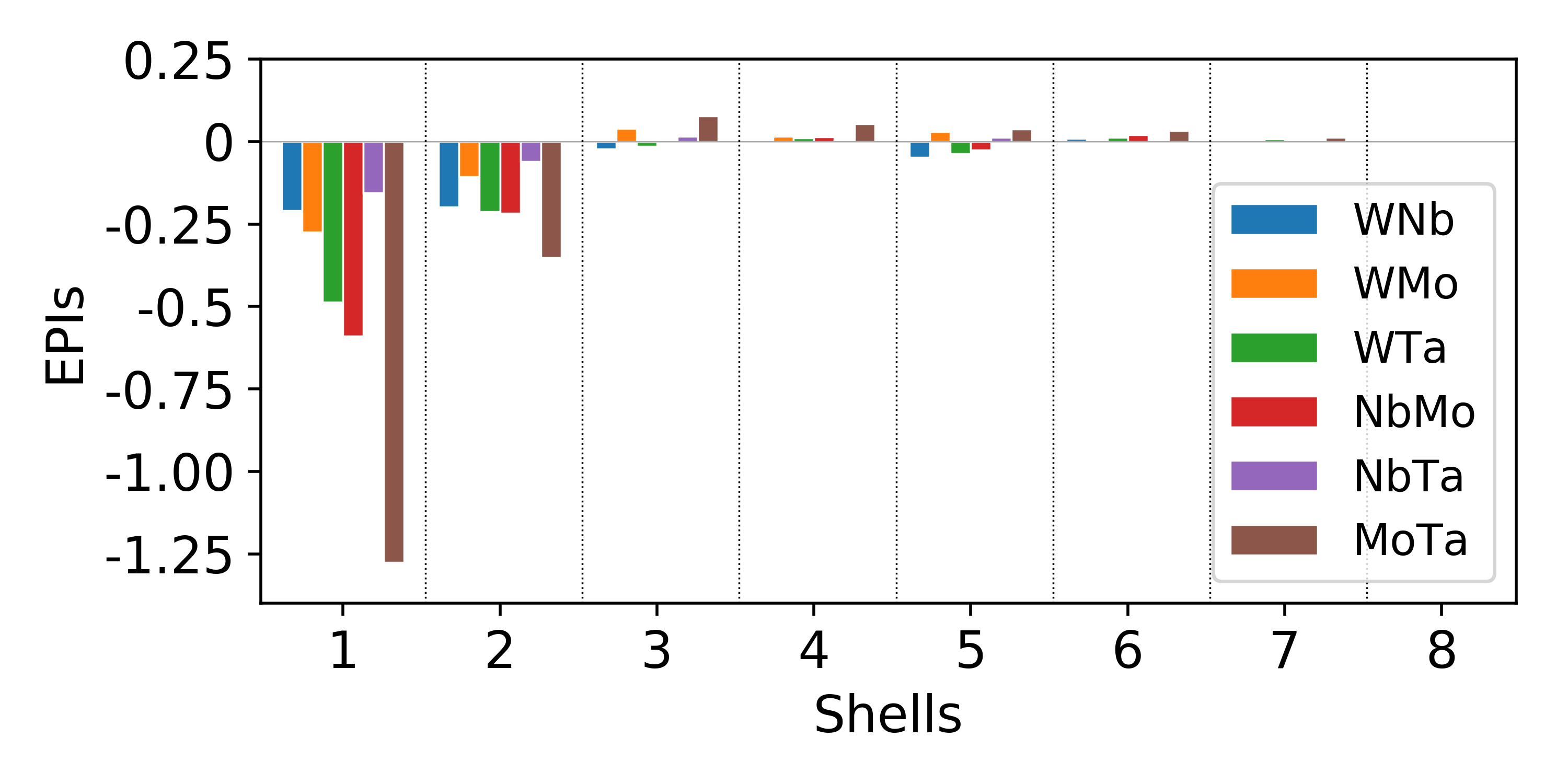}}
\subfigure[]{\includegraphics[width=0.45\textwidth]{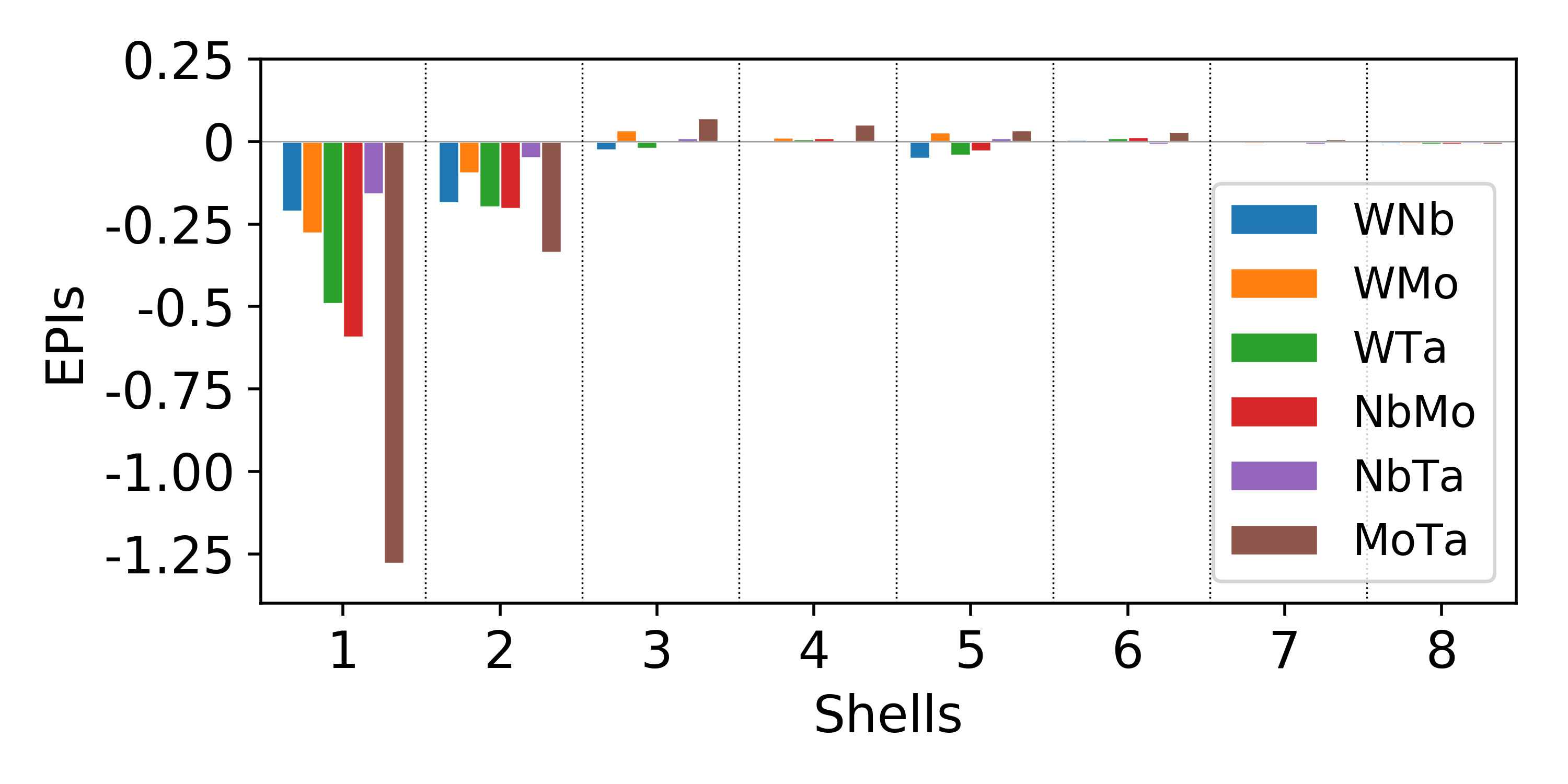}}
    \caption{Effective pair interactions (EPIs) results given different number of coordination shells from 1 to 8, which corresponds to the subfigures (a)-(h) respectively} \label{fig:f7}
\end{figure}

The impact of coordination shells can also be analyzed from the EPI parameters, as shown in Fig. \ref{fig:f7}. Firstly, it is easy to see that the EPI parameters corresponding to the first two shells play a dominant role, followed by these from 3rd to the 6th shells, while the EPIs of the 7th and 8th shells are negligible. This is in agreement with our previous observation of the fitting results. Second, the values of the EPI parameters are very consistent with respect to the number of shells considered in the models. Even for the most complicated 8 shell model (Fig. \ref{fig:f7}(h)), the EPIs of the 1st shell are still quite similar to that of the nearest-neighbor model (Fig. \ref{fig:f7}(a)), which demonstrates the advantage of using a physical quantity, i.e., EPI, to construct the effective Hamiltonian. Moreover, if one inspects the effect of elements in Fig. \ref{fig:f7}, it can be seen that the magnitude of the nearest neighbor EPIs are largest for MoTa, followed by NbMo and WTa, and are generally small for WMo, NbTa, and WNb. These phenomena can actually be easily understood by taking a look at the periodic table: Among the four elements, Mo and Ta are the most distinct in terms of electronegativity on the periodic table (along the northeastern direction), therefore tend to form a strong bond. NbMo and WTa are bonds between elements of the same period but neighboring VB and VIB families. WMo and NbTa are bonds between elements of the same family but neighboring periods, therefore are generally less favored, and WNb is just the opposite of MoTa.

\begin{figure}[!ht]    
    \centering
    {\includegraphics[width=0.55\textwidth]{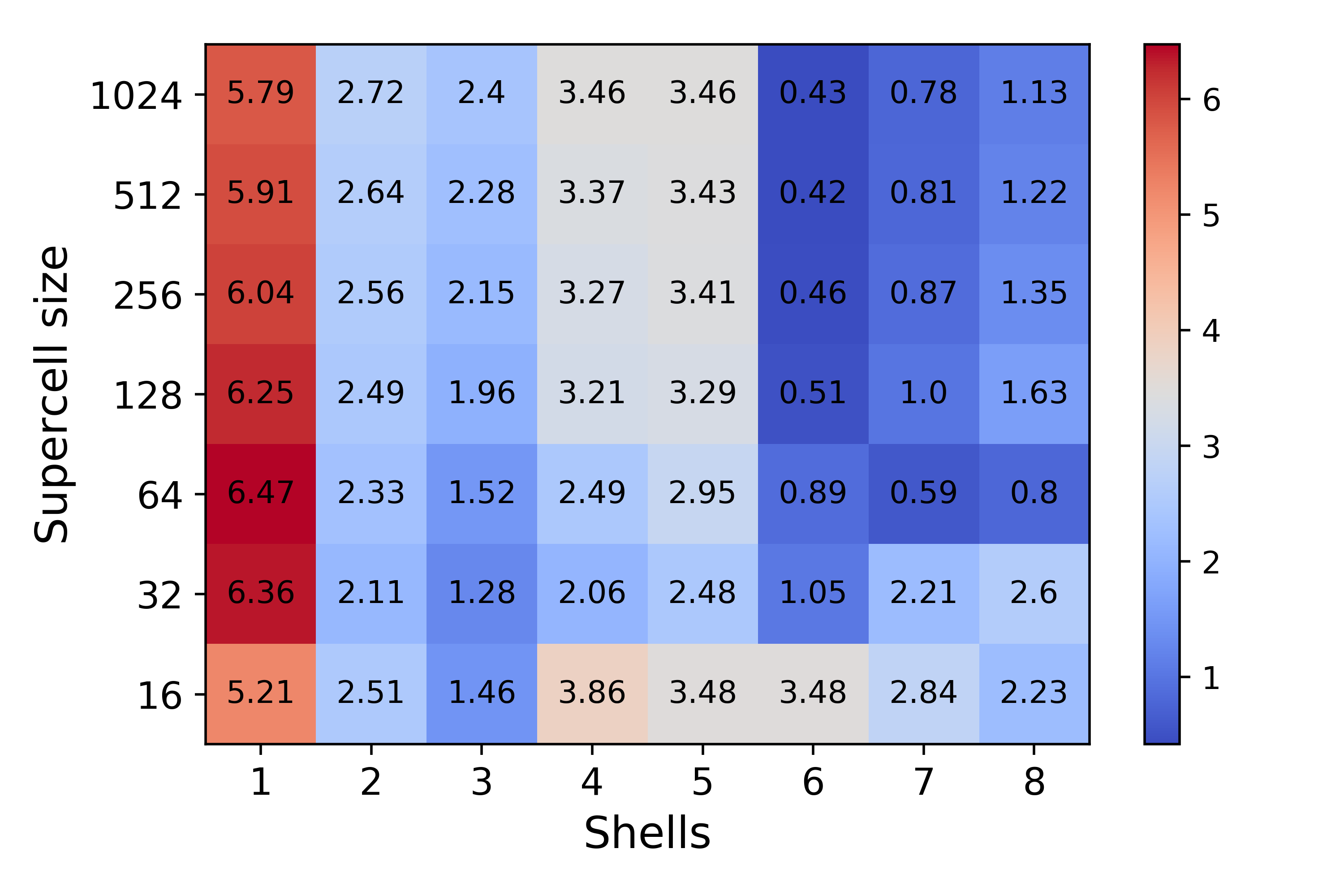}}
    \caption{Relationship heatmap between the number of coordination shells and and increased training data set (200, 400, 600, 800, 1000 and 1200) drawn from various supercell systems}  \label{fig:f8}
\end{figure}

A heat map summarizes the impact of training data and feature space on the prediction performance is shown in Fig. \ref{fig:f8}. It can be seen that the nearest-neighbor model will consistently produce an error of approximately 6 meV, no matter how many training data utilized. Such a large error is basically unacceptable for Monte Carlo simulations given that the total range of the configurational energy is only about 50 meV (see Fig. \ref{fig:f5}). On the other hand, using data from only small supercell also limits the accuracy of the model, and the data from at least 64-atom supercell are required to bring the testing error down to 1 meV. In addition, increasing the number of training data generally improve the testing results, just as expected. The best model is attained when the interaction within the 6th shells are considered, which also agrees with our previous observations.

\subsection{Neural networks}
\begin{figure}[!ht]    
    \centering
    \subfigure[]{\includegraphics[width=0.4\textwidth]{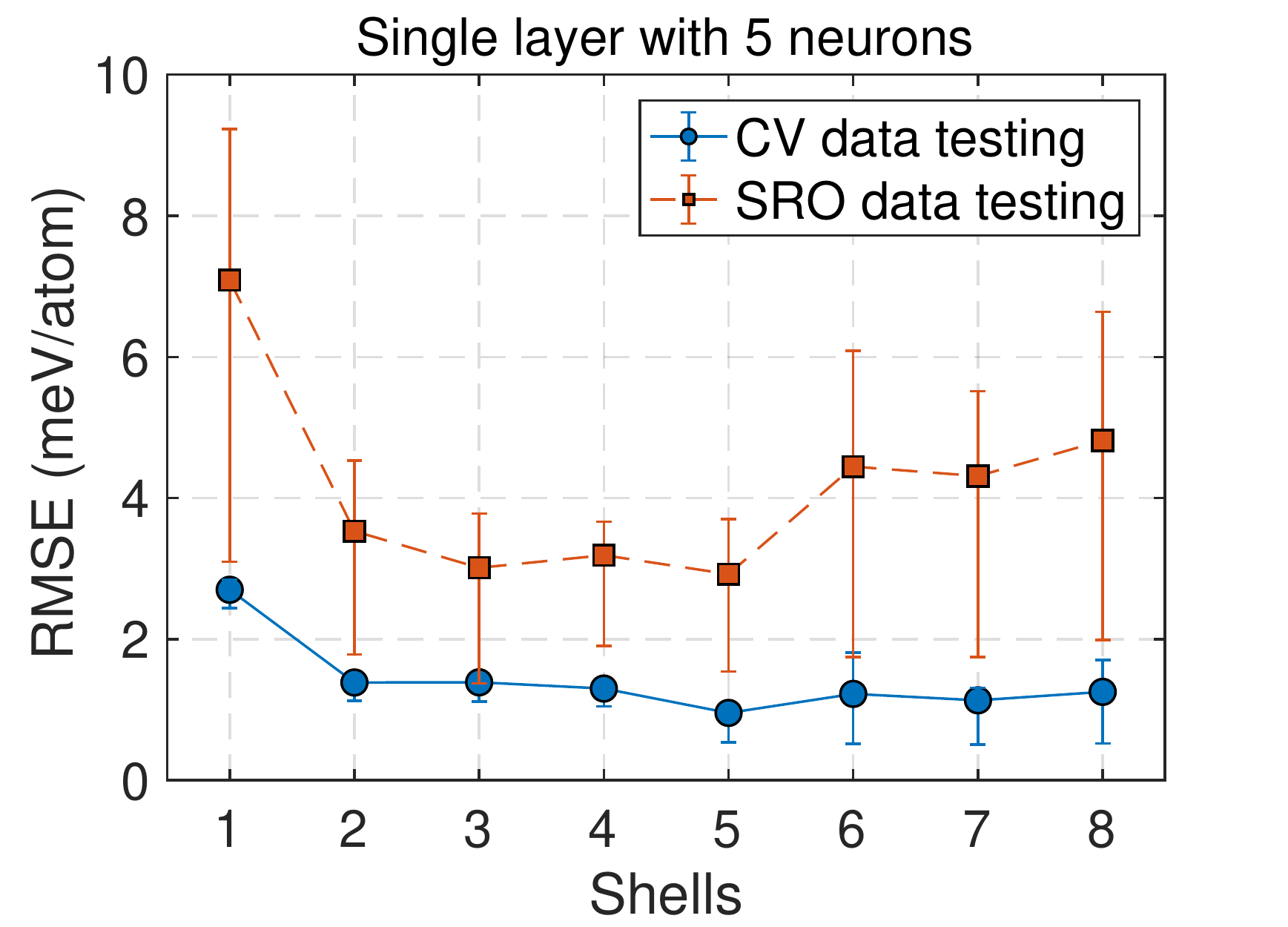}}
    \subfigure[]{\includegraphics[width=0.4\textwidth]{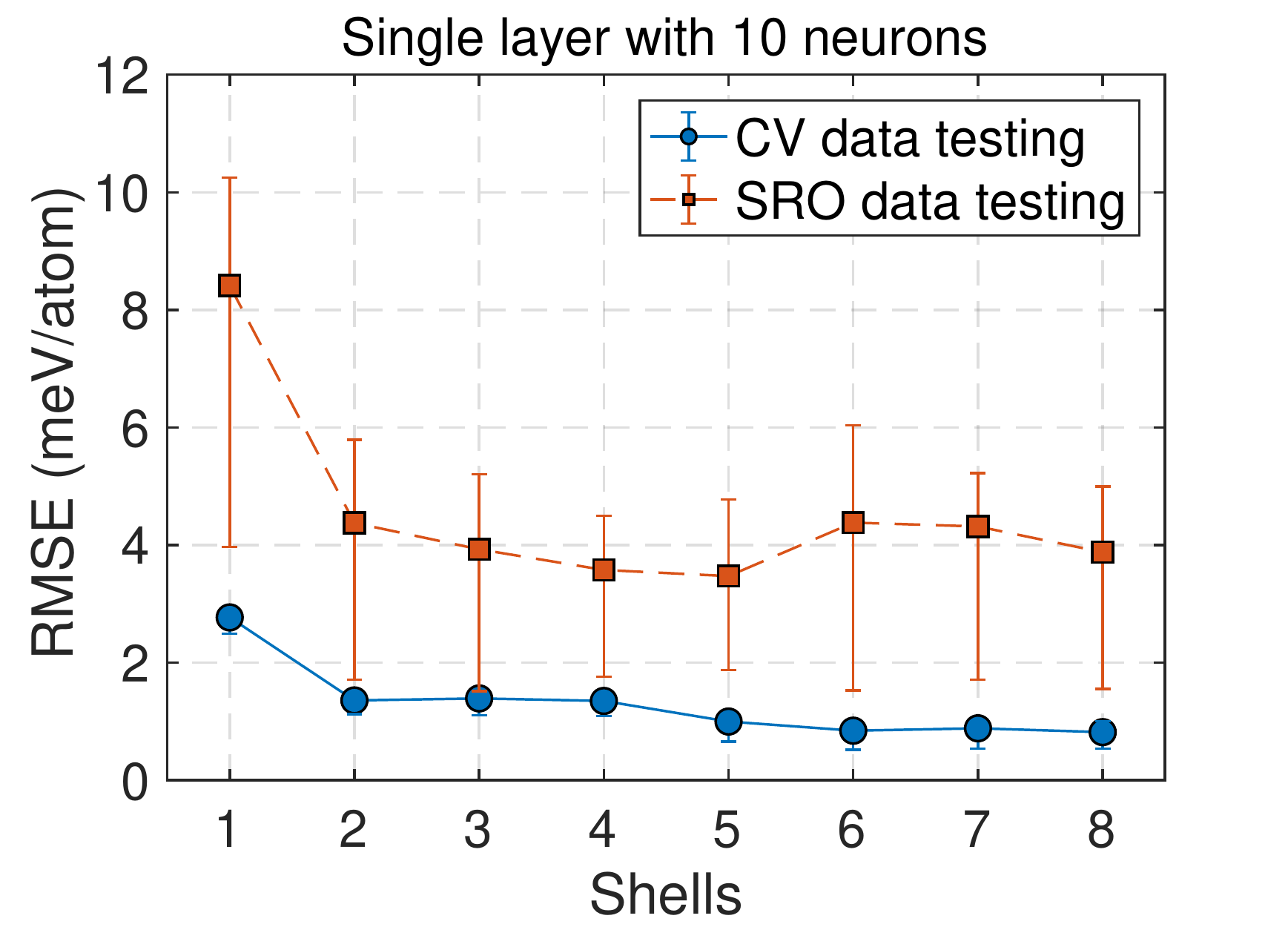}}
    \subfigure[]{\includegraphics[width=0.4\textwidth]{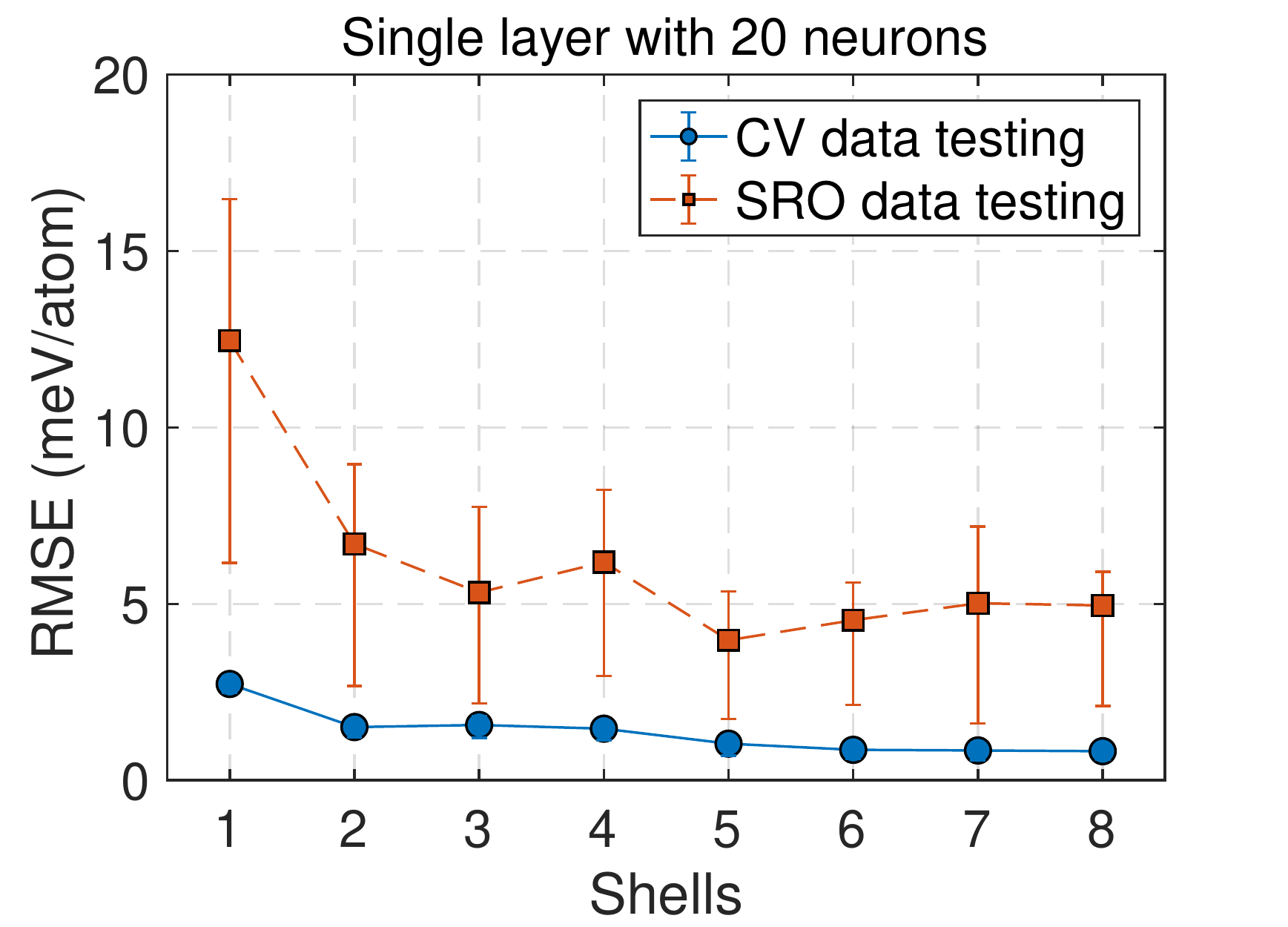}}
    \subfigure[]{\includegraphics[width=0.4\textwidth]{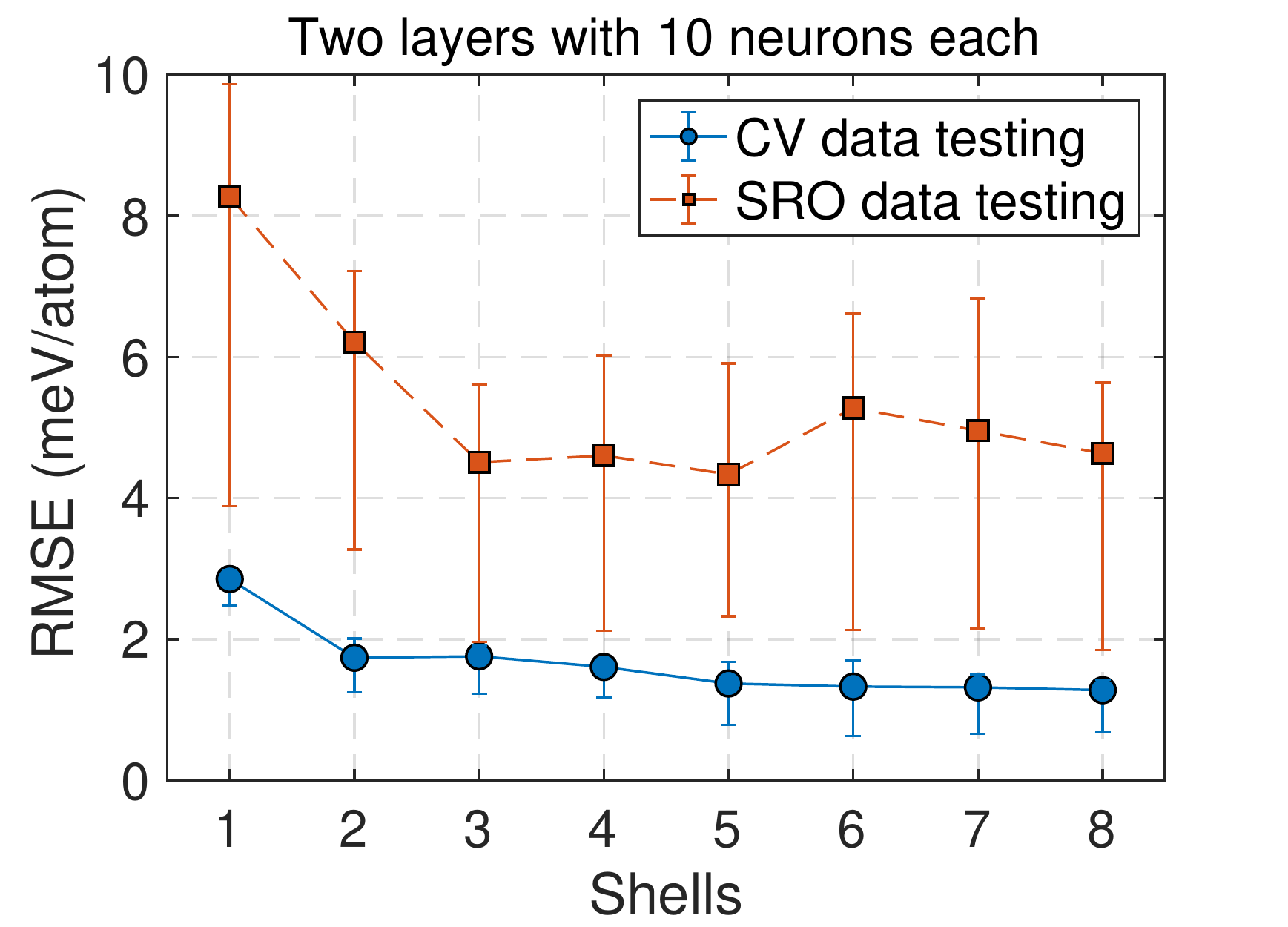}}
    \caption{Machine learning prediction using neutral networks (NN) with various structures, including (a) single layer with 5 neurons, (b) single layer with 10 neurons, (c) single layer with 20 neurons and (d) two layers with 10 neurons for each layer} \label{fig:f9}
\end{figure}

Neural networks are widely used to train models of large amount of features and capture complex relationship between the inputs and outputs. For a given atomic configuration $\sigma$, the loss function $\mathcal{L}$ in the neural network is defined in terms of the predicted energy $E(\sigma)$ and DFT calculated energy $E_{DFT}(\sigma)$ as
\begin{equation}
\mathcal{L} = \frac{1}{N}\sum_{\sigma}\left(E(\sigma)-E_{DFT}(\sigma) \right)^2. \label{Loss}
\end{equation}
Since the performance of neural networks depends on their structure, four different neural network structures are examined and the results are shown in Fig. \ref{fig:f9}. 
Testing of the neural networks is performed using both the SRO data and 5-fold cross validation. From Fig.  \ref{fig:f9}, we can easily observe that all four neural network models demonstrate a much smaller RMSE in cross validation than the SRO data testing, indicating the neural networks overfit the training data. The overfitting can also be seen from the large error bar of the SRO testing. 

While the neural networks demonstrate a strong tendency to overfit, their versatility still gives them great potential on the modelling of high entropy alloys, which grants more future investigation on the improvement. On one hand, the added complexity in the model naturally requires more data points due to the bias-variance tradeoff. On the other hand, the incompetence of neural networks in extrapolating also originates from the loss of physics, which is in stark contrast to the pair interaction model that is made up of only effective bonds and SRO parameters. Therefore, one strategy to improve the neural networks in the future might be taken physics into account when constructing the neural networks. 

\section{Summary}
The application of machine learning to the construction of effective Hamiltonian sheds new light on the traditional problem of thermodynamics in complex multicomponent systems. A lot of progress has been made, as demonstrated by the various modeling schemes proposed \cite{ML_Cluster,Korman_npj}. However, a key factor in any statistical modeling, representativeness of the data, has been largely missed. In this work, we propose a simple technique to obtain representative data: combining the DFT data calculated with different sizes of supercells. This method naturally incorporates chemical configurations of various short-range and long-range order, therefore provides a good sampling of the configuration space. We also propose to use effective pair interactions to construct the effective Hamiltonian of HEA systems, with the SRO parameters as features of the linear regression model. Applying this method to the prototypical NbMoTaW HEA, we find the trained effective Hamiltonian provides a very accurate description of the configurational energy, achieving an $R^2$ testing score of 0.997 and root mean square error of 0.43 meV. Moreover, we find the widely used cross-validation generally underestimate the testing error of a predictive model. Therefore, one should be very cautious when using it to evaluate the performance of models. By comparison, the testing with data set independent of the training one would give a much more reliable evaluation of the model’s accuracy. Finally, our results also highlight the importance of incorporating physics into the model, as demonstrated by the robustness of the pair interaction model and the tendency of overfitting in the neural networks.  

\section{Method}
The DFT data are calculated with the locally self-consistent multiple scattering (LSMS) method \cite{PhysRevLett.75.2867}, which is a real space implementation of the Korringa-Kohn-Rostoker (KKR) method that scales linearly with respect to the number atoms. The angular momentum cutoff in the LSMS method is set as 3, a local interaction zone of 59 atoms is used, and the scalar-relativistic equations are solved to properly treat the heavier elements in the system.

The testing data sets are generated with a simulated annealing algorithm. This algorithm initializes the configuration randomly at high temperature, then progressively decreases the temperature and updates the configuration with the Metropolis algorithm, in order to reach the state of desired SRO parameters. In practice, we set the nearest-neighbor SRO parameters, i.e., WNb, WMo, WTa, NbMo, NbTa, MoTa as $\alpha \times$(1, 1, -1, -1, 1, -1), where 200 values of $\alpha$ distribute uniformly between 0 and 1. When $\alpha=1$ the SRO parameters correspond to the ordered ground state of Nb–Mo–Ta–W–W–Ta–Mo–Nb sequence in \cite{Korman_npj}, and when $\alpha = 0$ it is just a completely random configuration. If one intends to make more inclusive tests, other values of SRO parameters at different shells can also be added into the testing sample. The size of the supercell is 1024 atoms. After the configurations are generated, they are feed into the LSMS method to calculate the total energy.

The scikit-learn package is used for both linear regression and neural networks. All the linear regression results presented in this article are obtained with ordinary least square. We also test the effects of regularization schemes using both Lasso and Ridge regression, which turn out to have a minor effect on the results. Training neural networks is achieved by minimizing the loss function in Eq. \eqref{Loss} using backpropagation algorithms. In this study, we use advanced quasi-Newton methods that adaptively update the learning rates for each weight parameter with an initial rate of $\lambda=0.001$. We further use mini-batch training with mini-batch size $m=100$, where the gradients are calculated over a subset of training data before updating the weights. 1000 epochs of batch training are run for the training data set.

\section{Acknowledgements}
X. L was supported by the U.S. Department of Energy, Office of Science, Basic Energy Sciences, Materials Science and Engineering Division. J. Z. was supported by the Department of Energy, Laboratory Directed Research and Development funding. This research used resources of the Oak Ridge Leadership Computing Facility, which is supported by the Office of Science of the U.S. Department of Energy under Contract No. DE-AC05-00OR22725. 

\section{Contributions}
X. L. conceived the project and calculated the DFT data. J. Z. performed machine learning training. X. L. and J. Z. analyzed the results and wrote the draft. Y. W. and M. E. maintain the LSMS code. All authors contribute to  the final manuscript.

\bibliographystyle{model1-num-names}
\bibliography{sample.bib}







\end{document}